\documentclass[12pt]{article}

\usepackage[T1]{fontenc}
\usepackage[utf8]{inputenc}
\usepackage[letterpaper,margin=1in]{geometry}
\usepackage{amsmath}

\usepackage{newtxtext,newtxmath}
\usepackage{graphicx}
\usepackage[version=4]{mhchem}
\usepackage[numbers,sort&compress]{natbib}
\usepackage[font=small,labelfont=bf]{caption}
\usepackage{url}
\usepackage[colorlinks,allcolors=blue]{hyperref}

\begin{document}

\begin{center}
{\LARGE\bfseries X-ray Driven Trihydrogen Formation on Silica Nanosurfaces\par}
\vspace{0.5em}
{\normalsize Samuel~Sahel-Schackis$^{1,2,3*}$,
Adam~Summers$^{2}$,
Ritika~Dagar$^{1,4}$,
Alexandra~Feinberg$^{1,2}$,
Martin~Grassl$^{1,2}$,
Simon~Dold$^{5}$,
Rebecca~Boll$^{5}$,
Yevheniy~Ovcharenko$^{5}$,
Chris~Aikens$^{6}$,
Cesar~Costa~Vera$^{7}$,
Alberto~De~Fanis$^{5}$,
Avijit~Duley$^{6}$,
Felix~Gerke$^{8}$,
Daniel~Jost$^{2,9}$,
Regina~Leiner$^{10}$,
Michael~Meyer$^{5}$,
Ilana~J.~P.~Molesky$^{11}$,
Razib~Obaid$^{1}$,
Jeffrey~Powell$^{12}$,
Nils~Rennhack$^{5}$,
Bj\"orn~Senfftleben$^{5,13}$,
Hendrik~Tackenberg$^{14}$,
Paul~Tuemmler$^{14}$,
Sergey~Usenko$^{5}$,
Christian~Peltz$^{14}$,
Thomas~Fennel$^{14}$,
Markus~Gallei$^{10,15}$,
Eckart~R\"uhl$^{8}$,
Artem~Rudenko$^{6}$,
Daniel~Rolles$^{6}$,
Thomas~Linker$^{1,2}$,
Matthias~F.~Kling$^{1,2,16*}$\par}
\vspace{0.3em}
{\footnotesize\setlength{\baselineskip}{1.15\baselineskip}
$^{1}$Stanford PULSE Institute, SLAC National Accelerator Laboratory, Menlo Park, CA, USA.\\
$^{2}$Linac Coherent Light Source, SLAC National Accelerator Laboratory, Menlo Park, CA, USA.\\
$^{3}$Department of Physics, Stanford University, Stanford, CA, USA.\\
$^{4}$Department of Physics, Indian Institute of Information Technology Bhopal, Bhopal, Madhya Pradesh, India.\\
$^{5}$Operations Division, European XFEL GmbH, Schenefeld, Germany.\\
$^{6}$J. R. Macdonald Laboratory, Department of Physics, Kansas State University, Manhattan, KS, USA.\\
$^{7}$Department of Physics, Escuela Polit\'ecnica Nacional, Quito, Ecuador.\\
$^{8}$Institute of Chemistry and Biochemistry, Freie Universit\"at Berlin, Berlin, Germany.\\
$^{9}$Stanford Institute for Material and Energy Science (SIMES), SLAC, Menlo Park, CA, USA.\\
$^{10}$Polymer Chemistry, Saarland University, Saarbr\"ucken, Germany.\\
$^{11}$Department of Chemistry, University of Nevada, Reno, Reno, NV, USA.\\
$^{12}$Institut national de la recherche scientifique, Montr\'eal, Qu\'ebec, Canada.\\
$^{13}$Laboratory for Solid State Physics, ETH Zurich, Zurich, Switzerland.\\
$^{14}$Institute of Physics, University of Rostock, Rostock, Germany.\\
$^{15}$Saarland Center for Energy Materials and Sustainability, Saarland University, Saarbr\"ucken, Germany.\\
$^{16}$Department of Applied Physics, Stanford University, Stanford, CA, USA.\\
\vspace{0.2em}
$^{*}$Corresponding authors: samss@slac.stanford.edu, kling@stanford.edu\par}
\end{center}

\vspace{0.2em}

\begin{abstract}
\noindent
The trihydrogen cation (\ce{H3+}) initiates the ion-molecule reactions that build molecular complexity in interstellar space. Whether its canonical formation reaction \ce{H2+ + H2 -> H3+ + H} proceeds on inorganic surfaces under radiation-driven ionization has remained untested. Here we drive \ce{H3+} formation on hydrated silica nanoparticles using intense 1.88 keV X-ray pulses, combining ion velocity map imaging, electron time-of-flight spectroscopy, and single-particle coherent diffractive imaging to resolve this chemistry on individual particles. The self-induced surface electric field on the V/nm scale drives interfacial charge transfer and water fragmentation. This field is the dominant parameter governing the relative yields of \ce{H+}, \ce{H2+}, and \ce{H3+} across particle size, composition, and aggregation. Density functional theory and nonadiabatic quantum molecular dynamics simulations trace this field-driven charge transfer, directly analogous to band bending at semiconductor photoelectrodes. These results establish surface-field-driven charge transfer as a unifying mechanism between radiation dominated astrophysical environments and field-driven surface catalysis.
\end{abstract}

\clearpage

Trihydrogen (\ce{H3+}) is not only the simplest polyatomic molecular ion, but also the most ubiquitous triatomic ion in the cosmos \cite{tennyson_spectroscopy_1995}. Characterized by its cyclic $D_{3h}$ structure, \ce{H3+} behaves as a Brønsted-Lowry acid \cite{lowry_uniqueness_1923} and serves as a universal proton donor in interstellar chemistry \cite{smith_ion_1992,smith_dissociative_1993}. It acts as both a catalyst and a critically important precursor in the interstellar medium (ISM), protonating ions, atoms, and molecules to drive the formation of dense molecular clouds and complex organic compounds, which are believed to play a vital role in the creation of life in the Universe \cite{mccall_detection_1998,mccall_enhanced_2003}. Since its discovery by Thomson in 1911 \cite{thomson_xix_1912} and mechanistic elucidation by Hogness and Lunn in 1925 \cite{hogness_ionization_1925}, the formation pathway
\begin{equation}
    \ce{H2+ + H2 -> H3+ + H} \label{trihydrogen_eq}
\end{equation}
\noindent has been recognized as the most important reaction in astrochemistry \cite{oka_interstellar_2013}. This highly exothermic ($1.7$~eV) proton-hop reaction with a high Langevin rate constant ($\approx 2 \times 10^{-9} \text{ cm}^3 \text{ s}^{-1}$) proceeds via cosmic ray ionization of molecular hydrogen followed by proton transfer \cite{oka_infrared_1993}.

Theoretical \cite{palaudoux_formation_2019} and laboratory studies have extensively investigated \ce{H3+} formation from organic precursors using electron impact \cite{burrows_studies_1979,sharma_determination_2006,kushawaha_fragmentation_2008}, highly charged ion (HCI) collisions \cite{de_formation_2006}, intense laser fields \cite{furukawa_ejection_2005,okino_coincidence_2006,kaziannis_ejection_2009,ekanayake_mechanisms_2017,ekanayake_h2_2018}, and single-photon ionization by VUV and soft X-ray photons \cite{ruhl_charge_1990,thissen_fragmentation_1994,eland_origin_1996,pilling_production_2007}. These mechanisms typically involve bond cleavage and bond formation, where doubly ionized molecules fragment into neutral \ce{H2} and charged residues, followed by roaming hydrogen abstraction. In every case the precursor is an isolated organic molecule in the gas phase. Recent strong-field laser studies have shown that \ce{H3+} can form from purely inorganic precursors involving water on nanoparticle surfaces \cite{alghabra_anomalous_2021, dagar_tracking_2024, dagar_trihydrogen_2026}, a route of direct relevance to the water-covered silicate grains of dense molecular clouds, where amorphous silicate cores acquire \ce{H2O}-dominated ice mantles and serve as catalytic surfaces for molecular formation \cite{boogert_observations_2015,mcclure_ice_2023}. However, those experiments \cite{alghabra_anomalous_2021, dagar_tracking_2024, dagar_trihydrogen_2026} were driven by tunneling and rescattering dynamics that have no counterpart in radiation dominated environments. Whether the canonical reaction proceeds on a water-covered silicate surface under the inner-shell ionization and Auger cascades that actually initiate this chemistry in space has therefore remained untested.

X-ray free-electron lasers deliver fluences far above those of interstellar environments, but the per-event physics is shared: each absorption creates a \ce{Si} 1s core hole whose decay, through any of several Auger and cascade pathways, ejects several electrons. This route is most relevant to X-ray dominated regions, where X-rays from protostars, other young stellar objects, and active galactic nuclei irradiate interstellar dust \cite{wolfire_photodissociation_2022}. On the nanometer-scale grains at the small end of the interstellar size distribution, which contribute the bulk of the total grain surface area \cite{mathis_size_1977,weingartner_dust_2001}, a single such event transiently charges the grain to surface fields approaching the V/nm scale (see Supplementary Information). The larger particles studied here reach that same field regime within a single pulse, through many absorption events. The number of absorption events therefore differs while the resulting surface field does not, and it is the field that drives the surface chemistry. Because each particle intercepts the beam at a different position within the near-Gaussian focus, successive shots sample different fluences and, with them, different surface charges. The field is set by the charge and the particle radius together, so particles of different size reach the same field at different fluences. Binning by field rather than fluence is therefore what makes the comparison across sizes, and with interstellar grains, meaningful. Fields of this magnitude are otherwise reached only at scanning tunneling microscope tips and in plasmonic hot spots, where they are known to steer surface reactivity, so the same charging that drives interstellar chemistry also places these particles in a regime of interest to catalysis.

\begin{figure}[!ht]
  \centering
  \includegraphics[width=0.77\textwidth]{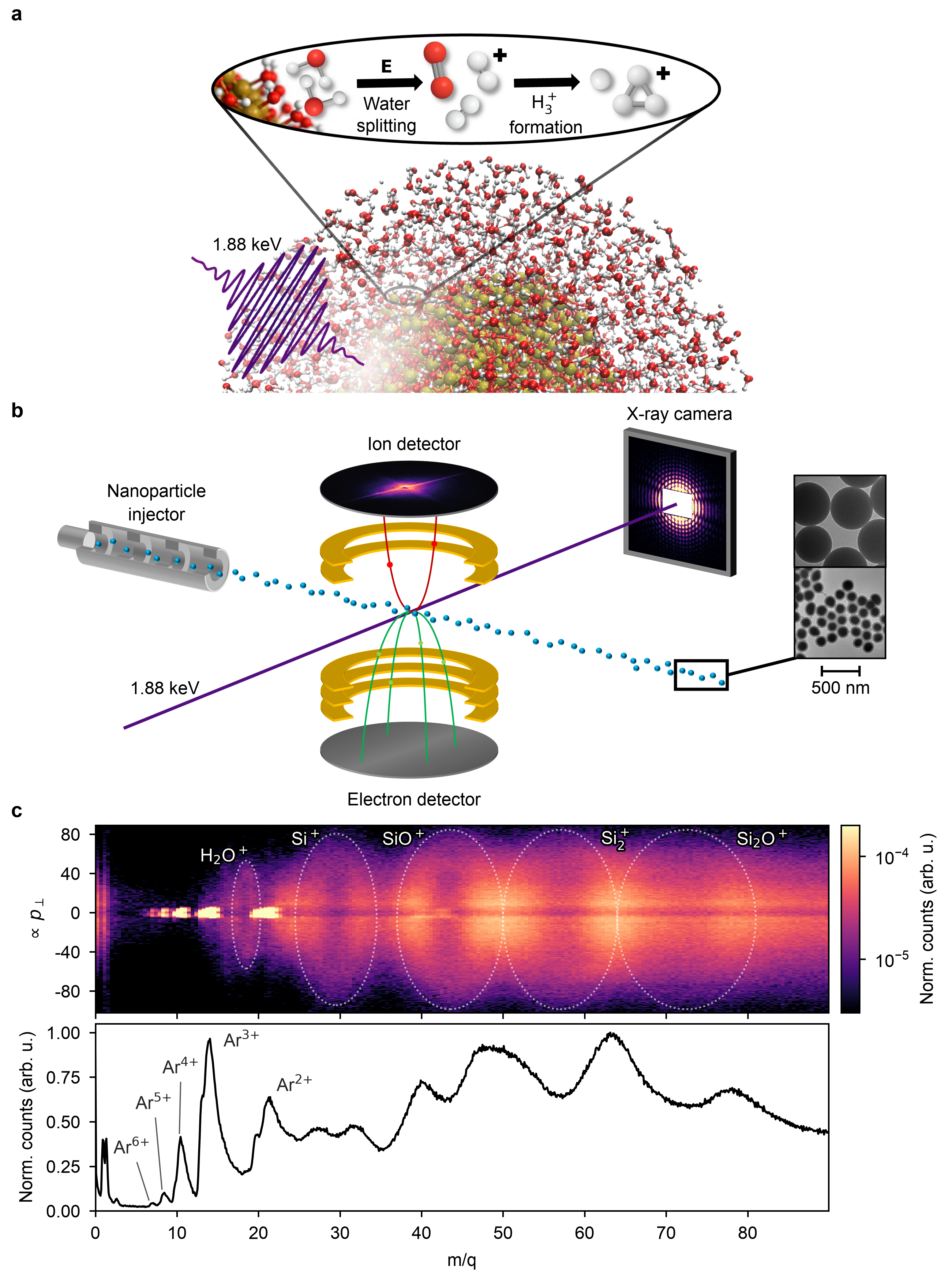}
  \caption{\textbf{X-ray charging of hydrated silica nanoparticles drives surface trihydrogen formation.}
    (a) A silica nanoparticle is ionized by a $1.88$~keV free-electron laser pulse, leading to surface electric fields on the V/nm scale. On the nanoparticle surface, water is ionized and fragments, including into \ce{H2} and \ce{H2+}. They give rise to the formation of \ce{H} and \ce{H3+}.
    (b) Experimental setup showing the integration of three complementary detection schemes: ion velocity map imaging (VMI), electron time-of-flight (eTOF) spectroscopy, and coherent diffractive imaging (CDI).
    (c) Transverse momentum versus mass-to-charge ratio $m/q$, and the corresponding mass spectrum, of the ionic species.}
  \label{fig1}
\end{figure}

Here we exploit this correspondence to drive astrophysically relevant surface chemistry on hydrated silica (\ce{SiO2}) nanoparticles using intense $1.88$~keV X-ray pulses, just above the \ce{Si} K-edge ($1.84$~keV). X-ray induced inner-shell ionization and Auger decay charge each nanoparticle, generating local electric fields on the V/nm scale that drive water splitting, silanol dissociation, and \ce{H3+} formation (Fig.~\ref{fig1}(a)). We resolve this chemistry shot-by-shot on individual nanoparticles and find that the self-induced surface electric field governs the relative yields of \ce{H+}, \ce{H2+}, and \ce{H3+}. This single quantity links the chemistry from water dissociation through to \ce{H3+} formation via the canonical reaction of Eq. \ref{trihydrogen_eq}. Together, these results reveal surface-field-mediated charge transfer as a unifying mechanism common to radiation dominated astrophysical environments and field-driven surface catalysis.

\section*{Multi-Modal Single-Particle Measurement}

Resolving the charge, fragmentation, and reaction dynamics driven by X-ray ionization requires measuring the ionic products, the electron yield, and the particle structure at the same time. Our multi-modal approach records ion velocity map imaging (VMI), electron time-of-flight (eTOF) spectroscopy, and coherent diffractive imaging (CDI) on the same nanoparticle, shot-by-shot, capturing the ionic fragments, the electron yield that fixes the final charge state, and the morphology that distinguishes monomers from dimers, respectively. This coincidence is what ties the ion and electron yields to the size, composition, and morphology of the specific particle that produced them.

In the experiment (Fig.~\ref{fig1}(b), Fig.~\ref{figS1}), isolated, water-covered nanoparticles are delivered into the interaction region of the SQS (Small Quantum Systems) instrument \cite{tschentscher_photon_2017} at the European XFEL \cite{decking_mhz-repetition-rate_2020}, where they are ionized by an intense $1.88$~keV X-ray beam just above the silicon (Si) K-edge ($1.84$~keV). The resulting 1s$^{-1}$ core holes decay within femtoseconds (Auger lifetimes of $\approx1.5$~fs for the 1s$^{-1}$ state and $\approx340$~as for the 2s$^{-1}$ cascade channel, Hartree-Fock estimates discussed in Methods), rapidly populating valence ionization states dominated by the O 2p orbitals. These ionization and cascade processes leave a positively charged nanoparticle whose confined surface charge generates strong electric fields on the V/nm scale. The fields drive surface fragmentation and chemical reactions, and the resulting ions are repelled from the surface by the net positive charge. Because each particle samples a different position within the Gaussian focus, the X-ray intensity, and with it the surface charge state, varies from shot to shot (Fig.~\ref{figS2}).

The ions leaving the surface are characterized by their mass-over-charge ratio ($m/q$) together with their transverse momentum (Fig.~\ref{fig1}(c)); spectrometer calibration and emission geometry are outlined in the Supplementary Information (Figs.~\ref{figS3} to \ref{figS6}). The carrier argon (Ar) gas is ionized to multiple charge states (\ce{Ar+}, \ce{Ar^2+}, \ce{Ar^3+}, \dots) without significant transverse momentum. In contrast, water ions (\ce{H2O+}), together with the fragments produced by decomposition of the silica nanoparticle, leave the surface with a range of kinetic energies that map onto radial positions on the detector. 

\section*{Surface Charge State}

The X-ray fluence experienced by each nanoparticle is monitored by the number of elastically scattered photons recorded on the pn-junction Charge Coupled Device (pnCCD) (see Supplementary Information) \cite{kuster_1-megapixel_2021}. For particles of $100-500$~nm diameter at $1.88$~keV the scattered intensity follows Mie scattering, with the total scattered signal increasing monotonically with the incident fluence on the particle. Independently, the total ion yield provides a second fluence monitor, correlated with the first (Fig.~\ref{figS7}), as such it is used as the hit intensity monitor in the following.

\begin{figure}[!htb]
  \centering
  \includegraphics[width=\textwidth]{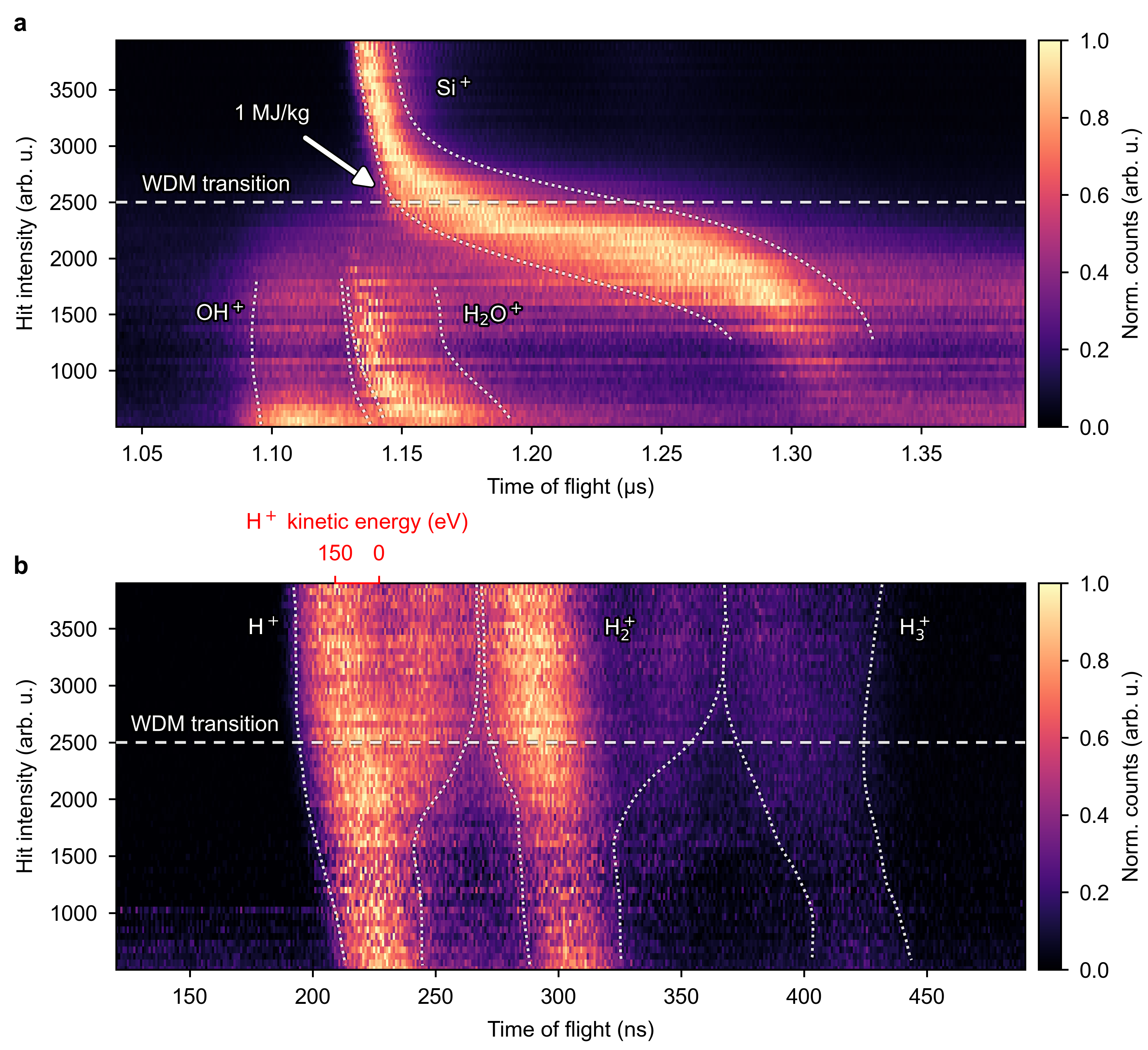}
  \caption{\textbf{Trihydrogen forms below the transition to warm dense matter.}
    (a) Hit intensity versus time-of-flight for ionized water (\ce{H2O+}) and its fragment (\ce{OH+}). At higher intensities the \ce{Si+} feature disperses to higher kinetic energy (shorter flight time), with a change of slope (arrow) at the warm dense matter (WDM) transition (deposited energy density $\approx 1$~MJ/kg).
    (b) Same intensity axis for the \ce{H+}, \ce{H2+} and \ce{H3+} features, from water dissociation and the canonical trihydrogen formation reaction. The \ce{H+} kinetic energy is inferred through SIMION simulations. \ce{H3+} forms below the WDM transition.}
  \label{fig2}
\end{figure}

Monitoring the silicon ion (\ce{Si+}) flight time as a function of this fluence, see Fig.~\ref{fig2}(a), reveals two distinct regimes separated by a change in the slope at a deposited energy density of $\approx 1$~MJ/kg, of order $0.1$~eV per atom, enough to heat the near-surface shell toward the \ce{SiO2} melting point (arrow Fig.~\ref{fig2}(a), full \ce{Si+} kinetic energy map in Fig.~\ref{figS8}). Below this threshold the nanoparticle remains molecularly bonded and the surface chemistry of interest proceeds. Above it, cooperative ionization drives the particle into the warm dense matter (WDM) regime, where the molecular framework breaks down and the chemistry gives way to plasma dynamics. Below the transition, \ce{Si+} leaves the surface of a charged particle with a kinetic energy fixed by the surface field, so its flight time tracks the charge state. Above the transition, \ce{Si+} is released from a disassembling, plasma-like particle rather than from the surface of an intact one, and the slope changes.

To quantify the surface field, two independent measures are combined. SIMION simulations \cite{dahl_simion_2000} convert the measured \ce{Si+} kinetic energy distribution ($0-500$~eV) for the $500$~nm particles (Fig.~\ref{figS8}) into surface fields of $0-2$~V/nm. Independently, the electron yield $N$ recorded by the eTOF provides the field on a sphere of radius $R$,
\begin{equation}
    E\left(R\right)=\frac{Ne}{4\pi\epsilon_0 R^2} \label{surface_field}
\end{equation}
\noindent treating the charged particle as a uniformly charged sphere and neglecting dielectric screening (see Supplementary Information, Fig.~\ref{figS9}). The two routes are independent and agree over their common range ($0-2$~V/nm). At higher fields, reached by smaller particles, the field follows from electron counting alone, whose inverse-square scaling with radius (Eq.~\ref{surface_field}) accounts for the order-of-magnitude rise from the $500$ to the $100$~nm particles. The electron yield measure provides the per-shot field determination used in the following sections.

\section*{Field-Driven Water Dissociation}

In the surface chemistry regime, conventional semiconductor photocatalysis splits water through band-gap excitation,
\begin{equation}
    \ce{2H2O +}h\nu\ce{ -> 2H2 + O2}
\end{equation}
\noindent but silica is a wide-bandgap insulator that is normally non-catalytic \cite{distefano_band_1971}. Here the activation comes from the surface field rather than from direct photoexcitation: multiscale simulations show that the V/nm fields on an ionized \ce{SiO2} surface drive rapid silanol dissociation and charge-transfer splitting of adsorbed water within $150$~fs \cite{linker_catalysis_2024}.

The underlying step is charge transfer. As the nanoparticle charges, valence holes left by the ionization cascade are driven by the surface field into the adsorbed water layer, where they fragment the water molecules. This is consistent with the surface-charge-driven weakening of surface \ce{O-H} bonds inferred from single-particle measurements and quantum dynamical simulations on silica nanoparticles \cite{dagar_tracking_2024}. The water cation \ce{H2O+} and its dissociation fragment \ce{OH+} are both clearly resolved (Fig.~\ref{fig2}(a)). The neutral \ce{H2} precursor is supplied by the same ionized-water chemistry. The water dimer radical cation dissociates predominantly by proton transfer, \ce{(H2O)2^{.+} -> H3O+ + OH^{.}} \cite{angel_dissociation_2001}. A competing channel, \ce{(H2O)2^{.+} -> H + OH^{.} + H2O+}, releases H atoms that recombine to \ce{H2}, a route established for ionized water at room temperature \cite{svoboda_reaction_2013,mi_formation_2022}. In radiolysis more broadly, the early-time \ce{H2} yield is attributed to electron-hole charge recombination \cite{sterniczuk_source_2016}, consistent with the field-driven charge transfer observed here. The \ce{H2+} precursor is seen directly in our spectra (Fig.~\ref{fig2}(b)). Therefore, both precursors \ce{H2} and \ce{H2+} of the canonical reaction (Eq.~\ref{trihydrogen_eq}) are present on the surface.

\section*{Field-Dependent Trihydrogen Formation}

The three hydrogen ion species are resolved in Fig.~\ref{fig2}(b), which maps \ce{H+}, \ce{H2+}, and \ce{H3+} against X-ray fluence. \ce{H3+} is clearly present below the WDM threshold, showing that trihydrogen forms in the surface chemistry regime. The following analysis is restricted to fluences below this transition, so that the measured yields reflect surface chemistry.

\begin{figure}[!htb]
  \centering
  \includegraphics[width=\textwidth]{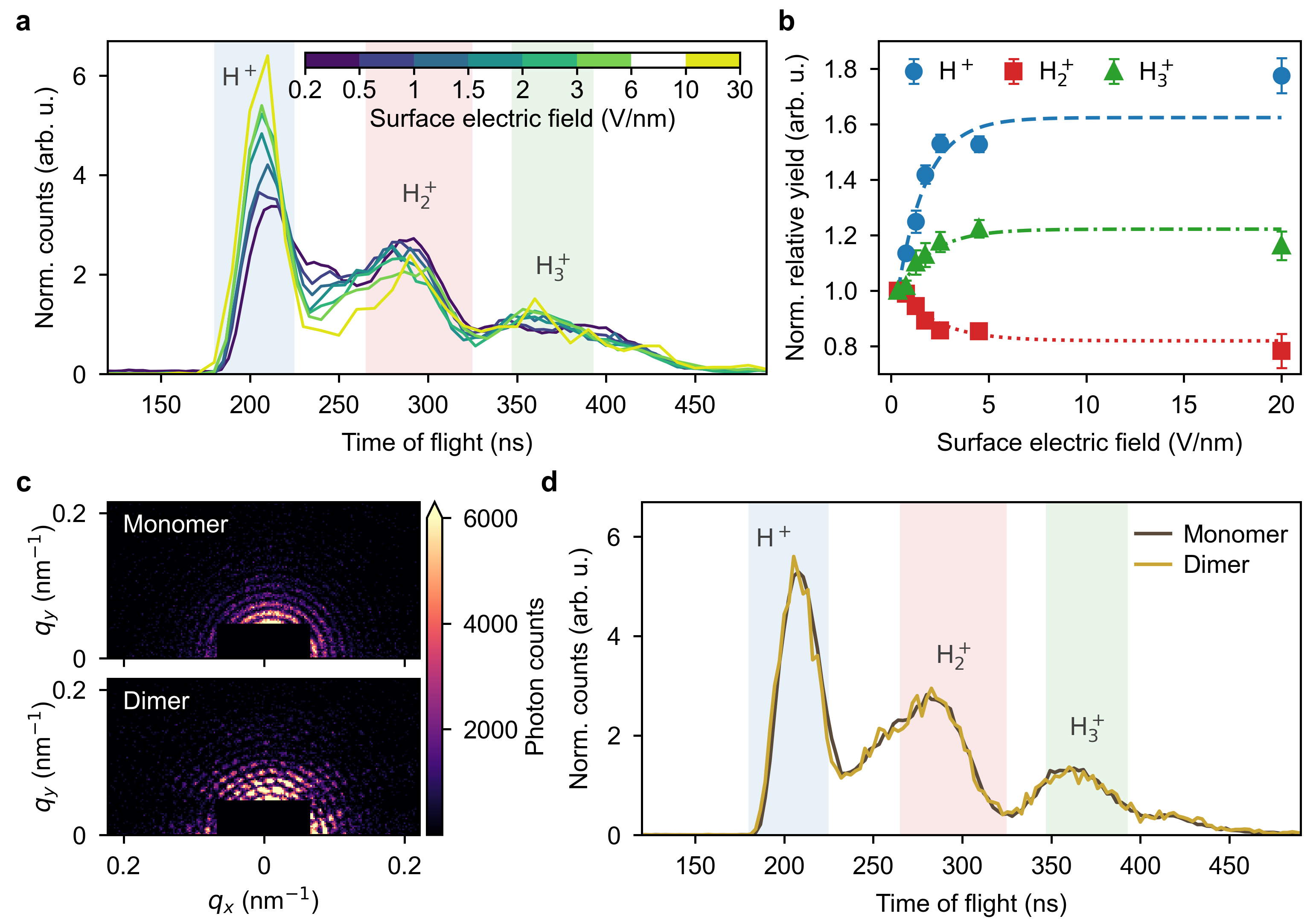}
  \caption{\textbf{The self-induced surface electric field governs the relative hydrogen ion yields.}
    (a) Time-of-flight spectra binned by surface electric field $E$, aggregated across all particle sizes and compositions within the surface chemistry regime.
    (b) Relative yields of \ce{H+}, \ce{H2+}, and \ce{H3+} versus $E$, with exponential fits. As the field increases, \ce{H+} and \ce{H3+} rise and \ce{H2+} declines. The field sets the balance among the three hydrogen ion products of surface water fragmentation.
    (c) Representative CDI patterns classified by a convolutional neural network as monomer (top) and dimer (bottom) for $300$~nm silica particles.
    (d) Time-of-flight spectra of monomer and dimer events selected at matched effective field ($2-4$~V/nm), where the dimer value includes the neighbor's contribution by superposition. The yields overlap independently of aggregation.}
  \label{fig3}
\end{figure}

The relative yields of the three hydrogen ions are set by the surface electric field. This is tested across silica particles of $500$, $300$, and $100$~nm diameter and core-shell \ce{Au}@\ce{SiO2} particles (a $100$~nm gold core within a $25$~nm silica shell). The latter are expected to enhance the surface yield because the high-$Z$ gold core has a substantially larger soft X-ray photoabsorption cross-section than silica at $1.88$~keV \cite{148746}, so it absorbs more strongly and drives a stronger photoionization and Auger cascade that charges the particle \cite{casta_electron_2015,lipp_quantifying_2022}. Because the surface field rises with decreasing particle radius (Eq.~\ref{surface_field}), the smaller particles reach the highest fields, so across all sizes and compositions the accessible field spans roughly two orders of magnitude (well beyond the $2$~V/nm of the $500$~nm particles alone). Binning the data by surface field (Fig.~\ref{fig3}(a)) collapses the yields onto common curves (Fig.~\ref{fig3}(b)) that depend only on the electric field $E$. \ce{H+} and \ce{H3+} rise with increasing field while \ce{H2+} falls, that is the field sets the balance among the three hydrogen ion products of surface water fragmentation. The collapse is insensitive to the exact fluence window below the transition, and the same field ordering persists in windows above it, into the plasma regime (Fig.~\ref{figS10}). The analysis is restricted to the surface chemistry regime, because only below the transition is the field acting on an intact hydrated surface, the configuration relevant to interstellar grains. Because Eq.~\ref{trihydrogen_eq} is barrier-free \cite{sanz-sanz_full_2013} and proceeds at the Langevin capture rate, the classical limit for a barrier-free ion-molecule reaction \cite{sanz-sanz_full_2013,savic_formation_2020}, this field dependence cannot originate in the reaction step. It must arise upstream, in the field-driven charge transfer and fragmentation that generate the \ce{H2} and \ce{H2+} precursors. Consistently, the ionized-water yield (\ce{H2O+}), a direct product of that upstream chemistry, shows the same field dependence, while the field-insensitive \ce{Ar^3+} gas reference stays flat (Fig.~\ref{figS11}). By contrast, the \ce{Si+} yield shows no monotonic field dependence, although its flight time shifts with field, as expected for an ion accelerated by the surface field. Since \ce{Si+} marks the fragmentation of the silica framework, its decoupling from the field shows that trihydrogen forms on the intact surface, before the particle disintegrates. An organic origin for the hydrogen ions is excluded, as no \ce{C+} or \ce{CH3+} signals are observed (Fig.~\ref{figS12}), ruling out residual ethoxy groups from the synthesis, or adventitious surface hydrocarbons, as the trihydrogen source.

Field control also holds across particle morphology. For the $300$~nm silica particles, CDI distinguishes monomers from dimers, where a dimer is any event involving two particles from contact to separation by up to $1$ $\mu$m, the resolution limit of the detector (Fig.~\ref{fig3}(c)). A convolutional neural network trained on the diffraction patterns, augmented with simulated dimer patterns to offset their low occurrence, classifies the two with an F1-score above $99\%$ (training details in SI, confusion matrix shown in Fig.~\ref{figS13}). For each event, the effective surface field is computed from the electron yield. In the dimer case, the superposition of the two particles' fields at the interparticle spacing measured from the diffraction fringes is accounted for (see Supplementary Information). Compared at matched effective field ($2$-$4$~V/nm, other ranges in Fig.~\ref{figS14}), where the dimer value already includes the neighbor's contribution by superposition, the monomer and dimer hydrogen ion TOF spectra coincide (Fig.~\ref{fig3}(d)). The chemistry thus depends only on the local surface field, whether self-induced (monomer) or augmented by a neighbor (dimer), and not on aggregation as such.

\section*{A Unified Field-Driven Mechanism}

\begin{figure}[!htb]
  \centering
  \includegraphics[width=\textwidth]{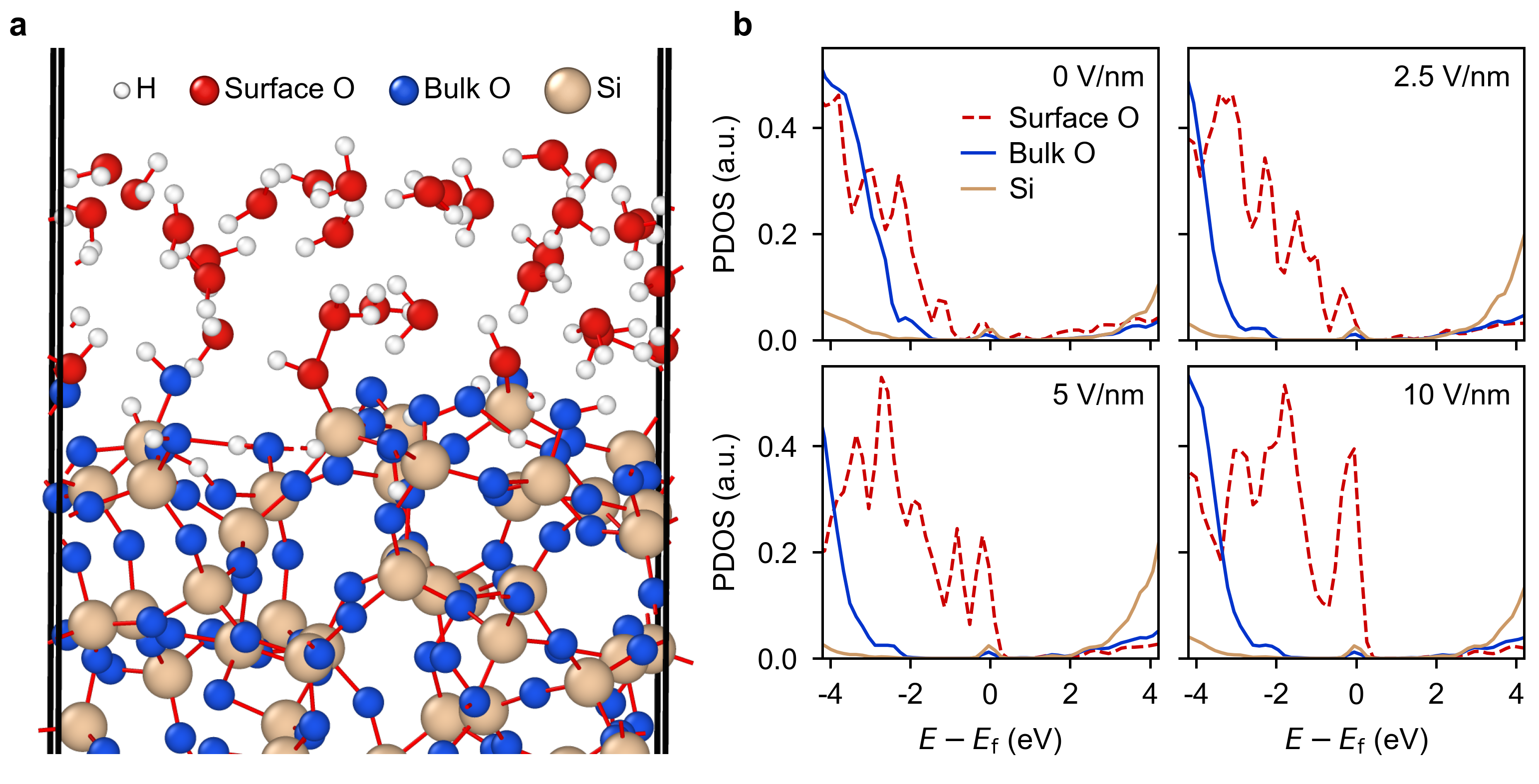}
  \caption{\textbf{Field-induced band tilting promotes hole transfer into the surface water layer.}
    (a) Nanometer-scale silica-water interface used in the first-principles simulations, representing the nanoparticle surface under the long-range charging field.
    (b) Atom-decomposed partial density of states (PDOS) of the interface under varying applied electric field. As the field increases, the water-derived O 2p states progressively shift into the intrinsic surface gap region, thereby increasing the driving force and hybridization for interfacial hole localization.}
  \label{fig4}
\end{figure}

What links these observations is a single field-driven mechanism, which is resolved with first-principles simulations. Following ref. \cite{linker_catalysis_2024}, the nanoparticle surface is represented as a nanometer scale silica water interface (Fig.~\ref{fig4}(a)) under an applied field that stands in for the particle's long-range charging field. The atom-decomposed partial density of states (PDOS) shows that, as the field increases, the water-derived O 2p states are progressively tilted into the intrinsic gap region associated with exposed Si surface sites (Fig.~\ref{fig4}(b)). This field-induced band tilting reduces the energy separation between the X-ray generated holes in the \ce{SiO2} surface states and the water acceptor states, increasing the driving force and hybridization for hole localization at the water layer.

Time-resolved nonadiabatic quantum molecular dynamics (NAQMD) confirm this picture. At $5$~V/nm, within the measured field range, the hole transfers from the silica into the water layer within $60$~fs (Fig.~\ref{figS15}), followed by water fragmentation driven by the surface field. In the simulations, double-hole transfer to the water layer yields a doubly charged water dimer whose Coulomb fragmentation, \ce{(H2O)2^2+ -> H2O+ + OH+ + H}, ejects the \ce{H2O+} and \ce{OH+} that leave the surface rapidly, whereas \ce{H3O+} remains bound within the hydrogen-bond network and is slower to desorb. This higher-charge channel coexists with the singly ionized one that supplies neutral \ce{H} (see ``Field-Driven Water Dissociation''), both populated by the Auger cascade. Simulations of ionized silica without surface water find \ce{H+} and \ce{OH+} ejection from silanol groups as the dominant channels \cite{linker_catalysis_2024}, consistent with the strong \ce{OH+} signal. The absence of an \ce{H3O+} signal (Fig.~\ref{fig2}(a)), together with the confinement of \ce{H3O+} in the simulations, points to its removal by dissociative electron recombination through the dominant channel, \ce{H3O+ + e- -> OH^{.} + H + H} \cite{neau_dissociative_2000}, releasing hydrogen atoms that recombine to \ce{H2}. The \ce{H2+} precursor is observed directly (Fig.~\ref{fig2}(b)) and is a known product of water-cluster ionization \cite{furuhama_reactions_2006}.

Both precursors of the canonical reaction are thus present on the surface. The same surface field that drives this water chemistry sets the \ce{H2}/\ce{H2+} precursor balance and hence the trihydrogen yield, so that one dominating parameter governs the chemistry from water dissociation through to \ce{H3+}. That this mechanism is independent of particle size, composition, and aggregation is exactly what the common curves of Fig.~\ref{fig3}(b) and the monomer/dimer comparison of Fig.~\ref{fig3}(d) establish. The picture is directly analogous to band bending-controlled charge separation at conventional semiconductor photoelectrodes, now operating at the V/nm fields self-induced by X-ray charging.

\section*{Outlook}

Trihydrogen forms on a hydrated silica surface through its canonical reaction, governed by a single self-induced quantity: the surface electric field set up by X-ray charging. That field sets the relative yields of \ce{H+}, \ce{H2+}, and \ce{H3+}, with monomer and dimer events coinciding at matched local field. DFT and NAQMD trace this control to field-driven charge transfer into the surface water layer, which supplies the \ce{H2} and \ce{H2+} precursors of the barrier-free reaction. A reaction central to interstellar chemistry, until now driven on surfaces only by strong-field lasers, is shown to proceed under radiation-driven ionization and to be controlled on that surface by the field. For astrochemistry, this identifies a new candidate route to \ce{H3+}, the ion whose abundance traces interstellar ionization rates. For catalysis, X-ray charging supplies a driving voltage with no electrodes and no applied bias. The resulting V/nm fields rival those at scanning tunneling microscope tips and in plasmonic hot spots, and at that strength could steer selectivity in reactions such as hydrogen evolution and \ce{CO2} reduction. Both directions build on the single-particle reactive scattering introduced here, which ties an individual particle's chemistry to its structure shot-by-shot and resolves the heterogeneity that ensemble measurements obscure.

\clearpage
\section*{Materials and Methods}

\subsection*{Experimental Setup}

The experiment was performed at the Nano-sized Quantum Systems (NQS) endstation of the SQS (Small Quantum Systems) instrument at the European XFEL \cite{decking_mhz-repetition-rate_2020}. Monodisperse silica (\ce{SiO2}) nanoparticles, with chosen diameters ranging from $100$ to $500$ nanometers, are dispersed in water (\ce{H2O}). An aerosol is formed by injecting argon (\ce{Ar}) gas into the suspension of nanoparticles in solution. The solvent is evaporated by a series of diffusion dryers, such that the gas-phase isolated nanoparticles are covered by a surface layer of water (\ce{H2O}) as they enter the vacuum interaction region where they intersect with the X-ray beam. The free-electron laser (FEL) was operated with a photon energy of $1.88$ keV, above the silicon (Si) K-edge of $1.84$ keV. The pulse energy upstream of the attenuator was $3$ mJ at a nominal pulse duration of $20$ fs. For the data in this publication, a gas attenuator (GAT) was set to $1\%$ nominal transmission. The focal spot size was not measured; from the beamline configuration it is estimated to lie in the $10-20$ $\mu$m range. Because this estimate spans a factor of four in area, and because the pulse duration and the beamline transmission downstream of the attenuator are nominal rather than measured, no peak intensity or on-target fluence is quoted. The shot-to-shot variation in the X-ray fluence experienced by each particle is instead monitored directly from the scattered photon and total ion yields (Fig.~\ref{figS7}). The European XFEL operates with an overall train repetition rate of $10$ Hz, whereas this experiment operated with $92$ pulses per train separated by $3.54462$ $\mu$s. This is necessary as heavy ions fly into the detector within microseconds to prevent wrap-around where those ions would be detected at early times.

The resultant photoelectrons and photo-ions are measured simultaneously using time-of-flight (TOF) and velocity map imaging (VMI) spectrometers (Fig.~\ref{figS1}) \cite{bromberger_shot-by-shot_2022}. The TOF follows a standard Wiley-McLaren design \cite{wiley_timeflight_1955}. The VMI spectrometer maps the velocity of the ions to different locations on the detector, independent of their initial positions. This is a result of the shape of the inhomogeneous fields, which form an Einzel lens \cite{eppink_velocity_1997}. Combining it with a Timepix3 camera, a timestamping camera, facilitates full 3D ion momentum imaging over time. The voltages of the conically shaped electrodes are set such that the Timepix VMI side is used to detect ions, while the TOF functions as an electron detector.

Meanwhile, a large-area, pn-junction Charge Coupled Device (pnCCD) detector capable of single-photon counting \cite{kuster_1-megapixel_2021}, is mounted in the forward scattering direction to monitor a uniform size distribution of the nanoparticles in the interaction region and provide single shot coherent diffractive imaging (CDI) measurements of our samples. CDI captures the far-field scattering pattern produced when a coherent X-ray beam interacts with a nanoparticle, enabling reconstruction of its real-space structure \cite{miao_extending_1999,miao_computational_2025}. In the case of spherical or near-spherical nanoparticles, this scattering produces Mie-type diffraction patterns characterized by concentric Airy rings. The number and spacing of these rings encode the particle’s size and ellipticity: smaller particles yield wider ring spacing, while anisotropic shapes distort the ring symmetry. Because the X-ray beam is larger than a single nanoparticle, both particles of a dimer are illuminated simultaneously, and their scattered fields interfere. The fringe modulation reveals both the individual particle sizes and their separation distance. Finally, the total scattered intensity in each shot reflects the local field intensity experienced by the nanoparticle, which varies depending on the particle’s exact position in the near-Gaussian beam focus (see Fig.~\ref{figS2}), thereby influencing its ionization level and resulting charge state.

\subsection*{Nanoparticle Synthesis}

Ammonia ($25\%$) was purchased from Fisher Scientific. Tetraethyl orthosilicate (TEOS, $\geq 99\%$) was purchased from Sigma Aldrich. Ethanol ($99.5\%$, denatured with $1\%$ MEK) was purchased from VWR Chemicals. All chemicals were used as received and Milli-grade water was used in all preparations. Monodisperse silica spheres of diameter $500$ and $300$ nm with a concentration of $10$ g/L in a solution of USP purified water were purchased from nanoComposix. Following the procedure for the stepwise growth of silica particles described by Graf \& van Blaaderen \cite{graf_metallodielectric_2002} using TEOS, ethanol, and ammonia, silica particle dispersions with a final solid content of $19$ g/L ($2.5$ wt\%) were prepared. Before use, the dispersion medium, containing ethanol and ammonia, was replaced with deionized water by centrifuging the synthesized particle dispersion at a nominal speed of $5000$ rpm for $5$ minutes. The supernatant was removed and replaced with an equivalent amount of deionized water. The new mixture of particles and water was re-dispersed. These preparation steps were repeated three times. The monodisperse spherical gold-silica core-shell nanoparticles with a concentration of $1$ g/L in a solution of USP purified water were purchased from nanoComposix.

For transmission electron microscopy (TEM) imaging, the particle dispersions were diluted and drop-cast on a carbon-coated copper grid. After the dispersion medium evaporated at ambient conditions for at least $12$ h, the TEM experiments were carried out using a JEOL JEM-2100 LaB$_6$ electron microscope (JEOL Ltd., Tokyo, Japan) at a nominal acceleration voltage of $200$ kV with a Gatan Orius SC100 CCD camera in bright-field mode. The particle sizes were analyzed using the software ImageJ. All the particles were analyzed via TEM, where the dried particles were imaged as perfect spheres (see Fig.~\ref{figS16}). The spherical silica particle diameters were measured to be $504 \pm 32$ nm, $312 \pm 14$ nm, and $126 \pm 7$ nm. The spherical gold-silica core-shell particles were measured to possess a gold core of $100 \pm 7$ nm, with the total core-shell being $147 \pm 9$ nm. They are respectively denoted $500$ nm, $300$ nm, $100$ nm \ce{SiO2} and $150$ nm core-shell \ce{Au}@\ce{SiO2} nanoparticles in the text. The measured diameters are used for calculations of the electric fields.

\subsection*{Data Analysis}

Per-shot X-ray fluence is monitored by the total scattered photon number and total ion yield, which are correlated (Fig.~\ref{figS7}). The surface electric field per shot is extracted from the eTOF electron yield via Eq. \ref{surface_field}, and cross-checked against SIMION simulations of the \ce{Si+} kinetic energy distribution ($0-500$ eV leads to $0-2$ V/nm for the $500$ nm particles). The two agree over their common range within the dielectric screening approximation discussed in the SI. The pooled field-dependence analysis (Fig.~\ref{fig3}(b)) uses an ion-yield window of $500-2000$ counts, below the warm dense matter transition for the larger particles and reaching it only for the smallest (see Supplementary Information); Fig.~\ref{figS10} shows the same redistribution for windows strictly below the transition ($500-1000$, $500-1500$), across it ($500-2500$, $500-3000$), and above it ($3000-4000$, $3000-5000$). The monomer/dimer comparison (Fig.~\ref{fig3}(d)) uses a separate, higher-intensity selection (stated in the SI), the larger ion counts reflecting the two-particle signal. For dimers identified by CDI, the per-event effective field includes the superposition of the two particles' fields at the diffraction-derived interparticle spacing (see Supplementary Information). Monomer/dimer classification uses a convolutional neural network trained on diffraction patterns with simulated dimer augmentation; architecture, training, and validation are in the SI.

\subsection*{Simulations}

Hartree-Fock calculations for atomic lifetimes were performed in the XATOM software using Hartree-Fock-Slater functional \cite{jurek_xmdyn_2016}. DFT simulations of the partial density states under different electric fields were performed utilizing the GPAW software \cite{mortensen_gpaw_2024}. Calculations are performed in a real-space basis on a finite difference grid of $0.2$ Å grid spacing within the projected augmented wave vector (PAW) method \cite{blochl_projector_1994}. The Perdew-Burke-Ernzerhof (PBE) version \cite{perdew_generalized_1996} of the generalized gradient approximation (GGA) was used and van der Waals corrections were employed utilizing the DFT-D scheme \cite{grimme_consistent_2010}. NAQMD simulations were performed in the QXMD software \cite{shimojo_large_2013,shimojo_qxmd_2019} using a plane wave basis with a $35$ Ry cutoff within the PAW method. NAQMD allows for dynamics of excited carriers and corresponding excited state forces to be modeled within the framework of time-dependent DFT. Excited state transitions are modeled within the fewest switch surface hopping method \cite{tully_molecular_1990}. For further details of the NAQMD algorithm we point readers to reference \cite{shimojo_large_2013}.

\clearpage
\section*{Acknowledgments}
We acknowledge European XFEL in Schenefeld, Germany, for provision of X-ray free-electron laser beamtime at Scientific Instrument SQS (Small Quantum Systems) under proposal number $3408$ and would like to thank the staff for their assistance. The authors thank Marcus Koch (Institute for New Materials, Saarbrücken) for supporting the transmission electron microscopy measurements. We thank Aiichiro Nakano (University of Southern California) and Fuyuki Shimojo (Kumamoto University) for access to the NAQMD software.  
\\\\
\noindent\textbf{Funding:}
This work was supported by the U.S. Department of Energy, Office of Science, Basic Energy Sciences, under Contract No. DE-SC0063. Simulations were performed at the Stanford Sherlock cluster and the National Energy Research Scientific Computing Center (NERSC), a Department of Energy User Facility using NERSC award BES-ERCAP 0035639. S.S.S. and T.L. acknowledge support by the U.S. Department of Energy, Office of Science, Office of Basic Energy Sciences under Contract No. DE-AC02-76SF00515. R.D. acknowledges support from INSPIRE Faculty Fellowship Scheme (DST/INSPIRE/IFFCALL-2024/IFA24-PH324). C.A., A.R., and D.R. are funded through the Chemical Sciences, Geosciences, and Biosciences Division, Office of Basic Energy Sciences, Office of Science, U.S. Department of Energy under grant no. DE-FG02-86ER13491. A.D. was supported by grant no. PHYS-2409365 from the National Science Foundation. E.R. was supported by Freie Universität Berlin. M.Ga. and R.L. are supported by the German Research Foundation (GRK 3082: Engineering covalent bonds in molecules and materials, Ec = m2), project number 534930008.
\\\\
\noindent\textbf{Author contributions:}
S.S.S. analyzed the data. S.S.S. and T.L. performed the simulations. A.S., A.R., D.R. and M.F.K. conceived the study. C.A., A.D., A.R. and D.R. provided the nanoparticle injector, A.S., R.D., C.A., A.D., J.P. and A.R. operated it. A.S., R.D., A.F., M.Gr., S.D., R.B., Y.O., C.A., C.C.V., A.D.F., A.D., F.G., R.L., M.M., I.J.P.M., R.O., J.P., N.R., B.S., H.T., P.T., S.U., C.P., E.R., A.R. and M.F.K. performed the experiments. R.L. and M.Ga. synthesized and characterized the nanoparticles. All authors discussed and interpreted the results, and contributed to the final version of the manuscript.
\\\\
\noindent\textbf{Competing interests:}
There are no competing interests to declare.
\\\\
\noindent\textbf{Data availability:}
Data recorded for the experiment at the European XFEL is available at \url{https://doi.org/10.22003/XFEL.EU-DATA-003408-00} with a current embargo, which will be lifted ahead of publication.
\\\\
\noindent\textbf{Code availability:}
The code used to analyze the data of the experiment can be found at \url{https://github.com/samsahsch/EuXFEL_nano_analysis}

\clearpage
{\small
\bibliographystyle{sn-nature}
\bibliography{references}
}

\clearpage
\appendix
\renewcommand{\thefigure}{S\arabic{figure}}
\setcounter{figure}{0}
\renewcommand{\theequation}{S\arabic{equation}}
\setcounter{equation}{0}

\section{Supplementary Text}

\subsection{Single-Event Charging of Interstellar Grains}

The correspondence between this experiment and X-ray dominated interstellar
environments rests on the surface field reached per ionization event rather than on
the absolute fluence. For a uniformly charged sphere the surface field follows
\begin{equation*}
    E = \frac{Ne}{4\pi\varepsilon_0 R^2} = 1.44 \frac{N}{R^2} \text{ V/nm,}
\end{equation*}
with $R$ in nm and $N$ the net number of ejected electrons.

A single inner-shell ionization above the Si K-edge produces a $1s^{-1}$ core hole
whose Auger cascade ejects of order two to four electrons; after partial recapture by
the charging grain the net charge per event is $N \approx 1-4$. On a grain of
radius $R \approx 1-3$ nm this gives $E \approx 0.2-6$ V/nm, the same
regime sampled by the collectively charged nanoparticles studied here. Because the
field scales as $N/R^2$, larger grains reach this regime through the accumulation of
several events rather than a single one.

Such nanometer-scale grains dominate the interstellar dust surface area: under a
standard size distribution $n(a)\propto a^{-3.5}$ \cite{mathis_size_1977}, the
differential surface area $a^2 n(a)\,\mathrm{d}a \propto a^{-1.5}\,\mathrm{d}a$ is
weighted toward the smallest grains, so the per-event field regime relevant to surface
chemistry is realized on the grains that carry most of the available surface.

The transient charge persists far longer than the chemistry it drives. The
field-driven water fragmentation and trihydrogen formation proceed within
$\sim 150$ femtoseconds to picoseconds in the simulations, whereas neutralization of an isolated
grain by electron recombination occurs on vastly longer timescales. The surface field
is therefore effectively static over the duration of the chemistry.

\subsection{Time-of-Flight Mass Spectrometry Calibration}

In the absence of a nanoparticle hit, the background consists of the argon (\ce{Ar}) carrier gas ionized to multiple charge states. The ionization of the noble gas produces no recoil, so spatially selecting the center of the VMI detector on these background pulses (Fig.~\ref{figS4}(a)) isolates the argon series in the time-of-flight spectrum. From longer to shorter flight times the ions \ce{Ar+}, \ce{Ar^2+}, \ce{Ar^3+}, \ce{Ar^4+}, \ce{Ar^5+}, and \ce{Ar^6+} are identified (Fig.~\ref{figS4}(b)) and assigned to their known mass-over-charge ratios $m/q = 40$, $20$, $40/3$, $40/4=10$, $40/5=8$, and $40/6$.

For an electrostatic spectrometer the flight time is related to the mass-over-charge ratio by \cite{mamyrin_time--flight_2001}
\begin{equation}
    \mathrm{tof} = t_0 + k\sqrt{m/q}, \label{eq:tofcal}
\end{equation}
where $k$ is set by the extraction field and flight geometry and $t_0$ is the fixed instrumental offset between the ionization trigger and the start of the drift region. A linear fit of the argon series to $\sqrt{m/q}$ with the exponent held at its physical value of $2$ then determines $t_0$ and $k$ for each of the two voltage settings used in this work:
\begin{enumerate}
    \item Voltage Setting 1 (Fig.~\ref{fig1}(c)): $t_0 = 7.45$~ns, $k = 2.062\times10^{-7}$~s.
    \item Voltage Setting 2 (all remaining figures): $t_0 = 6.82$~ns, $k = 2.450\times10^{-7}$~s.
\end{enumerate}

Fixing the exponent is justified rather than assumed: unconstrained fits return exponents of $2.013$ and $2.018$ across the two voltage settings, statistically indistinguishable from $2$ given the finite width of the peaks, and this small excess arises entirely from absorbing the nonzero $t_0$ into the power law when the offset is omitted. With $t_0$ included explicitly, the two-parameter form of Eq.~\ref{eq:tofcal} reproduces the argon series with a root-mean-square mass error of $\approx 0.06\%$, comparable to the centroid uncertainty of the peaks themselves, so the more complex three-parameter fit offers no improvement and the squared relation is adopted throughout.

\subsection{SIMION Simulations}

\begin{figure}[!htb]
  \centering
  \includegraphics[width=\textwidth]{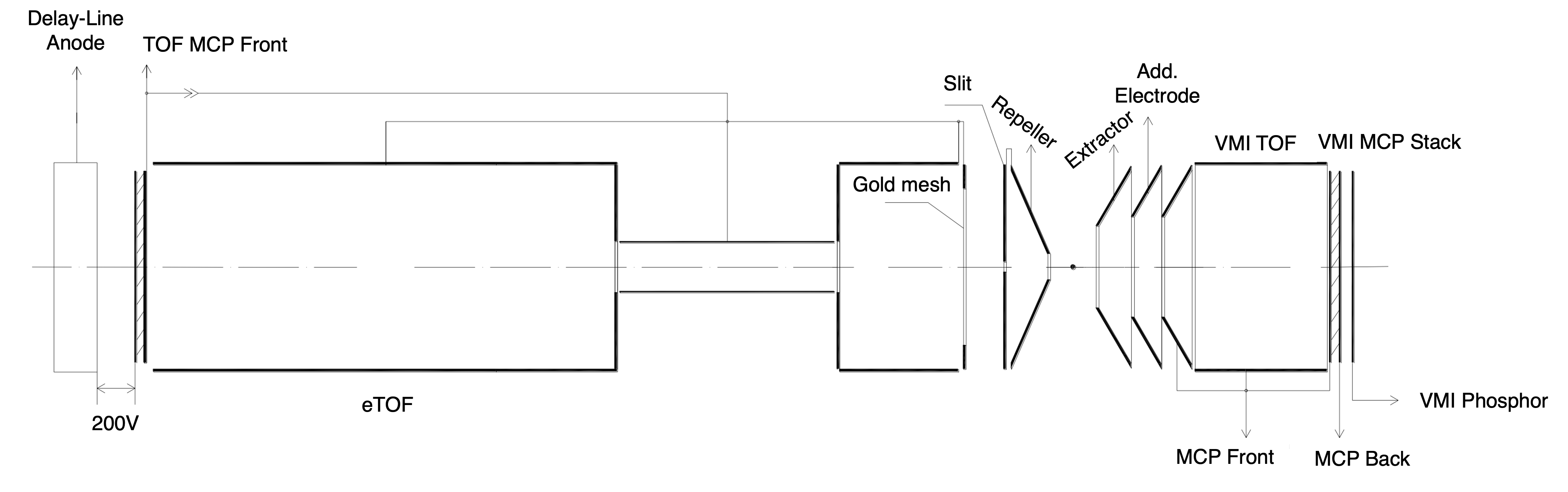}
  \caption{Sketch of the Nano-sized Quantum Systems (NQS) detector comprising both an ion Velocity Map Imaging (VMI) spectrometer and an electron time-of-flight (eTOF) spectrometer. Both are designed with conically shaped electrodes to minimize the shadow on the photon detector. The ion VMI is combined with a Timepix3 camera, a timestamping camera, recording ion flight times up to $3.54462$ $\mu$s. The eTOF follows a standard Wiley-McLaren design \cite{wiley_timeflight_1955}.}
  \label{figS1}
\end{figure}

To quantitatively relate the measured time‐of‐flight (TOF) and the spatial ion distributions on the Velocity Map Imaging (VMI) detector to the initial kinetic energies of the ejected ions, the electrostatic ion optics of the Nano‐sized Quantum Systems (NQS) endstation were modeled in SIMION 8.1 \cite{dahl_simion_2000}. The electrode geometry was reproduced from the NQS detector configuration (Fig.~\ref{figS1}). Simulations were used to determine both the relevant extraction voltages and the kinetic energies of the ejected ions based on their time-of-flight.

The electrode geometry was modeled according to the NQS detector configuration (Fig.~\ref{figS1}), using the repeller, extractor, and additional focusing electrodes, as well as the drift tube. A Gaussian ionization volume was defined at the laser and X-ray focus, with a transverse source size of $25$ $\mu$m perpendicular to the laser axis and the longitudinal interaction length being $3$ mm. The initial kinetic energy distribution of ions was modeled as a Gaussian. For each central kinetic energy value, $300$ trajectories were launched isotropically ($4\pi$ solid angle) for each ion mass. These simulations fixed the extraction voltages and provided the TOF to kinetic energy mapping used to convert the measured ion features into kinetic energies, as applied to the \ce{Si+} and hydrogen ion channels in Fig.~\ref{fig2}(a) and Fig.~\ref{fig2}(b).

A separate set of simulations established the calibration between the radial impact position on the VMI phosphor screen and the initial kinetic energy of the ions. By flying monoenergetic protons with initial kinetic energies between $1-100$ eV, a near-perfect power-law dependence between position and energy was found:
$$p = 5.69\,E^{0.5024}, \quad \text{and equivalently,} \quad E = 0.03188\,p^{1.986},$$
where $p$ is the radial position in mm and $E$ the kinetic energy in eV. The correlation exhibited an $R^2$ value of $0.99999$, indicating that the detector response is nearly ideal and square-root proportional, as expected for electrostatic focusing in VMI geometries.

\subsection{Separation of Reaction Mechanisms}

The transverse momentum of each ion is obtained from a spatial cutout on the VMI detector orthogonal to the X-ray beam, which is incident from left to right in the detector view (Fig.~\ref{figS3}). Two distinct reaction mechanisms are present in the data. The primary mechanism analyzed here produces ion emission that is uniform in all directions. A second mechanism arises from the argon carrier gas, whose ions are accelerated by the surface field and react with hydrogen ion species in the vicinity of the nanoparticle \cite{johnsen_chapter_2010,michaelsen_imaging_2017}; this channel is detected downstream of the interaction region and will be treated in a separate publication. To analyze the primary mechanism without imposing any momentum-based discrimination, a little under half of the detector is filtered out: in Fig.~\ref{figS5} the two bright spots along the X-ray direction mark the argon-related channel, and only data to the left of the red separation line are used.

\subsection{Ion Emission Geometry}
 
After ionization and photoelectron emission, each nanoparticle becomes a positively charged sphere (Fig.~\ref{figS6}). Ions produced at the surface, such as \ce{H+} from surface chemistry or \ce{Si+} when the framework fragments, are accelerated by the resulting field. Depending on their initial direction, ions either fly directly toward the detector, are ejected away from it and turned back by the applied extraction field so that they arrive later, or are ejected sideways with kinetic energy high enough to escape detection. This geometry is why the observed features appear as rings rather than filled disks in Fig.~\ref{fig1}(c).

\subsection{Fluence Monitor}
 
\begin{figure}[!htb]
  \centering
  \includegraphics[width=0.7\textwidth]{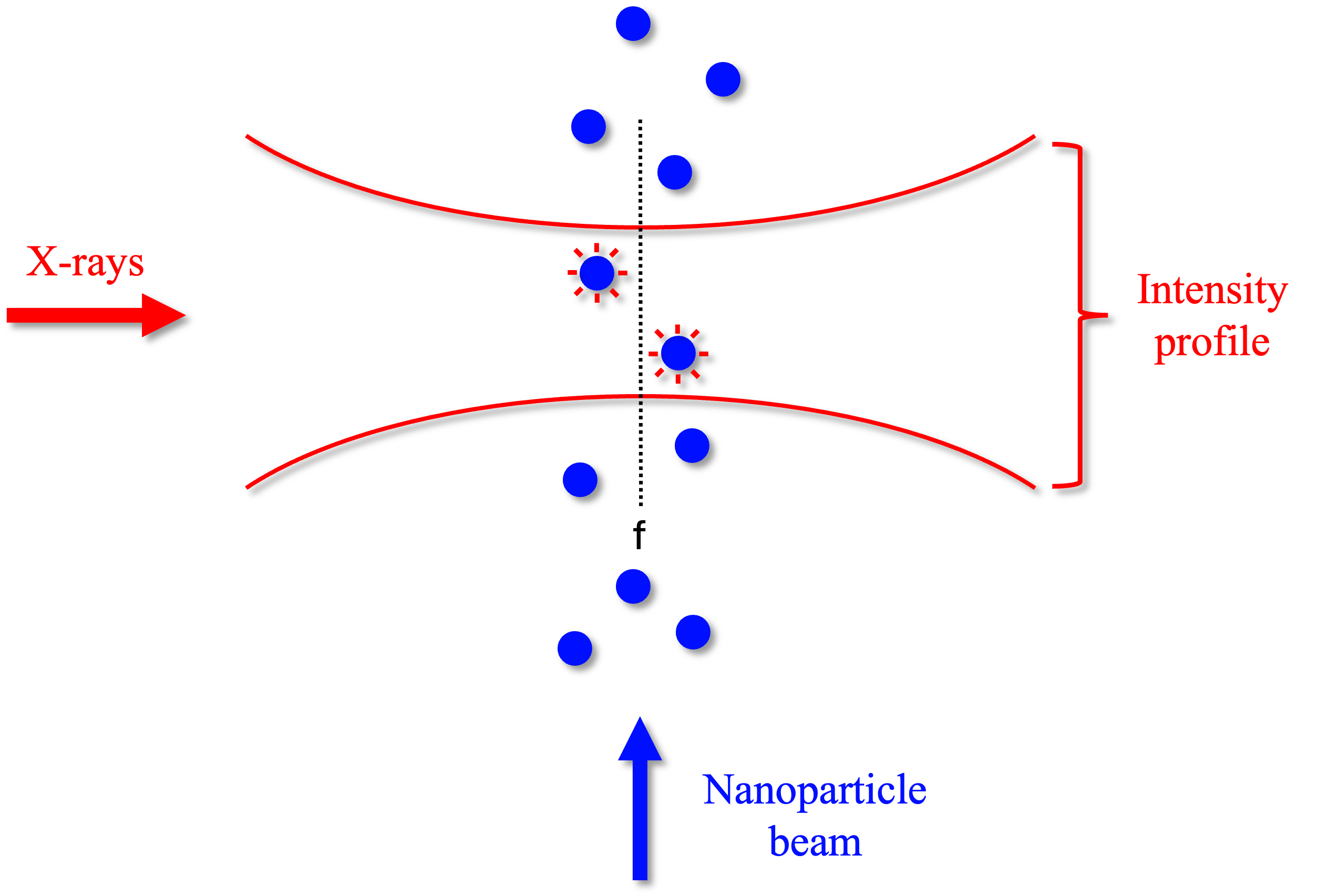}
  \caption{The ionization intensity the nanoparticles experience depends on their position within the Gaussian intensity profile of the X-ray beam.}
  \label{figS2}
\end{figure}

\begin{figure}[!htb]
  \centering
  \includegraphics[width=0.5\textwidth]{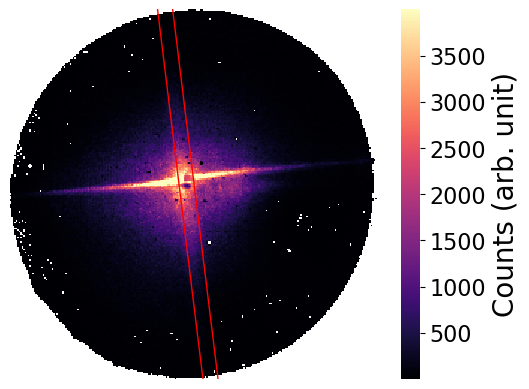}
  \caption{Spatial cutout on the ion VMI detector along the interaction region orthogonally to the X-ray beam direction, which in the figure is incident from left to right.}
  \label{figS3}
\end{figure}

\begin{figure}[!htb]
  \centering
  \includegraphics[width=\textwidth]{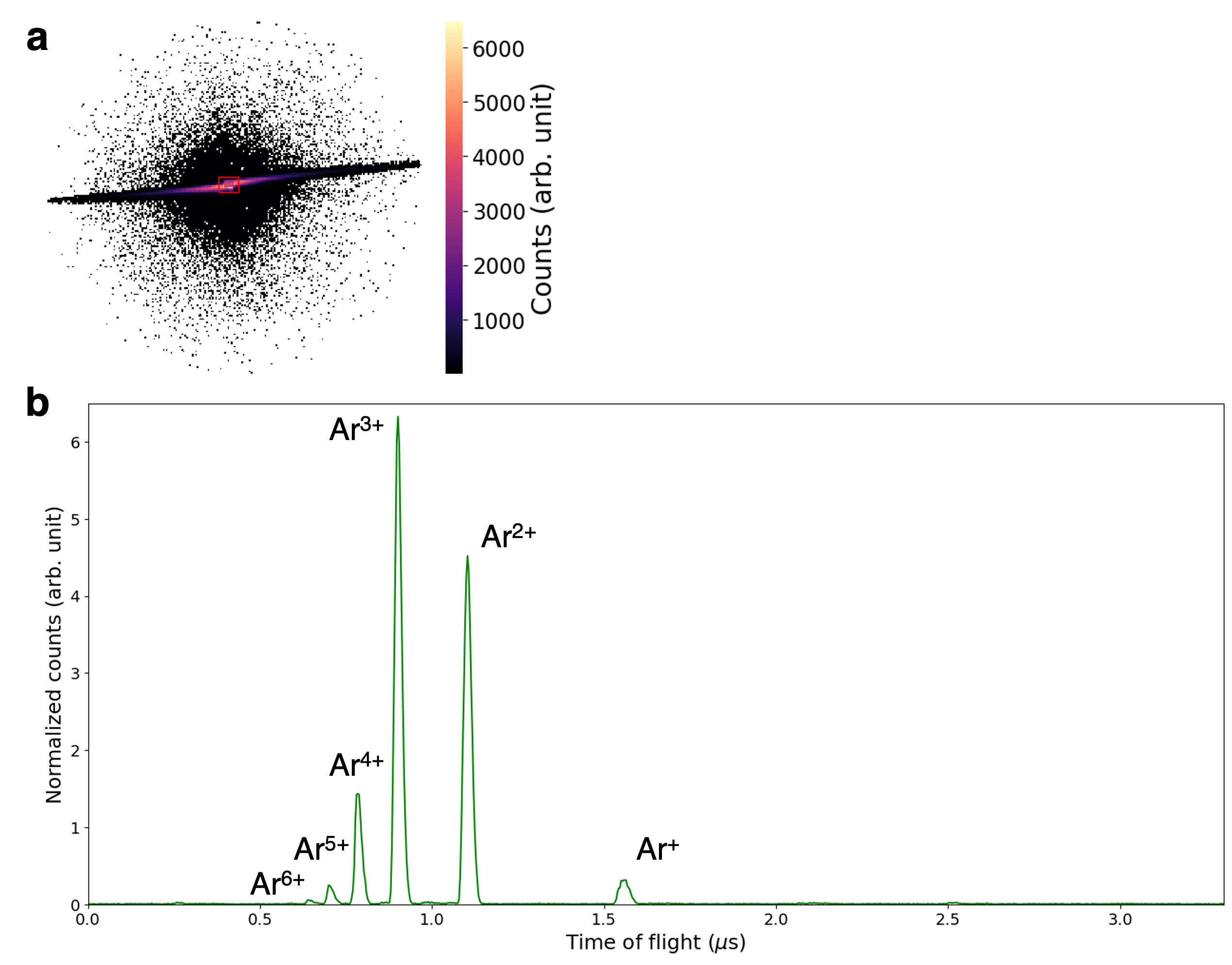}
  \caption{(a) Background hits on the ion VMI detector screened by selecting pulses with a low number of ion hits. The spatial cutout of the middle section selects ionized argon without momentum from the nanoparticle interaction region. (b) The time-of-flight spectrum of the downselected ions clearly showing the argon series peaks corresponding to argon ionized to multiple charged states (in order \ce{Ar^6+}, \ce{Ar^5+}, \ce{Ar^4+}, \ce{Ar^3+}, \ce{Ar^2+}, and \ce{Ar+}).}
  \label{figS4}
\end{figure}

\begin{figure}[!htb]
  \centering
  \includegraphics[width=0.5\textwidth]{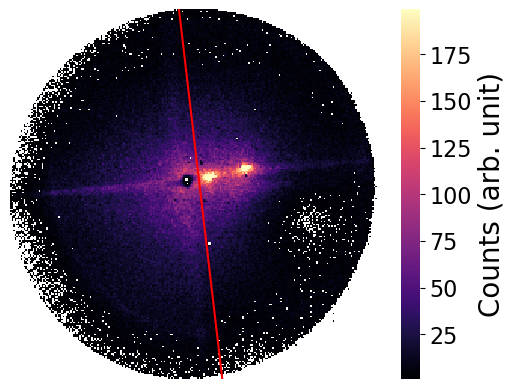}
  \caption{Spatial view of the ion VMI detector, with the X-ray beam direction incident from left to right. In the present publication, only data to the left of the red line is used, this is done to separate from a second reaction mechanism characterized by the two bright spots along the X-ray direction to the right of the red line.}
  \label{figS5}
\end{figure}

\begin{figure}[!htb]
  \centering
  \includegraphics[width=0.5\textwidth]{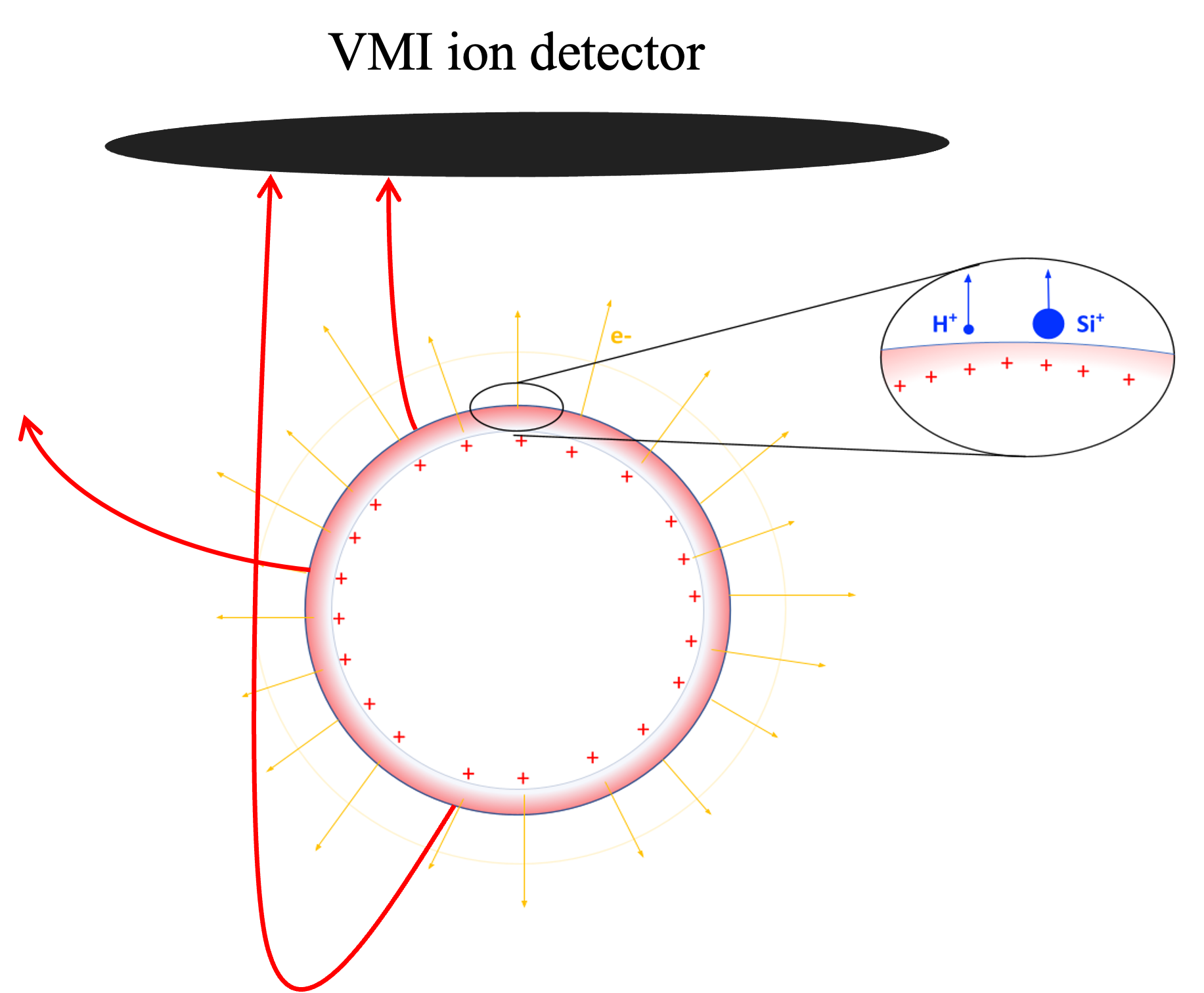}
  \caption{Scheme illustrating a positively charged sphere representing the nanoparticle after ionization and photoelectron emission. Catalytic reactions on the surface produce ions (e.g., \ce{H+}, or \ce{Si+} when the particle fragments), which are imparted kinetic energy by the surrounding electric field. The ions fly either towards the detector, away from the detector and turn around due to applied voltage thus arriving later, or sideways with high enough kinetic energy such that they are not detected.}
  \label{figS6}
\end{figure}

The X-ray fluence experienced by each nanoparticle varies shot-to-shot depending on the position of the nanoparticle within the Gaussian beam profile of the X-rays (Fig.~\ref{figS2}). The hit intensity can be inferred from the number of elastically scattered photons through Rayleigh scattering theory, 
$$P_\text{scattered} = \langle S_\text{incident}\rangle\,\Sigma_\text{scattering},$$
where $P_\text{scattered}$ is the scattered power, $\langle S_\text{incident}\rangle$ the incident irradiance, and $\Sigma_\text{scattering}$ the scattering cross-section. The number of scattered photons recorded by the CDI detector and the total ion yield recorded by the VMI are correlated (Fig.~\ref{figS7}), so either quantity serves as a per-shot monitor of the absorbed fluence. Each dot corresponds to one hit, accumulated over $500$, $300$, and $100$ nm \ce{SiO2} and $150$ nm core-shell \ce{Au}@\ce{SiO2} nanoparticles.

\subsection{Energy Density}

The energy density reached at the slope change of the \ce{Si+} feature (arrow in Fig.~\ref{fig2}(a); full kinetic-energy map in Fig.~\ref{figS8}) is estimated from the \ce{Si+} kinetic energy. The feature corresponds to the silicon ion (\ce{Si+}) channel emitted towards the detector. The SIMION trajectory simulations calibrated to the measured TOF spectra place this feature at $\approx 500$ eV, corresponding to a surface potential of $500$ V for a singly charged silicon ion on a $500$ nm diameter particle.

The number of absorbed photons is estimated from the total charge buildup required to reach this potential:
\begin{equation*}
    Q = 4\pi\varepsilon_0 R V_\text{surf},
\end{equation*}
which yields $N_+ = Q/e \approx 8.7\times10^4$ charges.

Assuming an effective emission yield of two electrons per absorbed photon, consistent with the photo-Auger decay cascade in \ce{SiO2} above the Si K-edge ($1.88$ keV) \cite{van_riessen_auger-photoelectron_2007}, the corresponding number of absorbed photons is $N_\text{abs} \approx 4.3\times10^4$.

The total deposited energy is therefore
\begin{align*}
    E_\text{abs} &= N_\text{abs}\,E_\gamma \approx 4.3\times10^4 \times 1.88\,\text{keV}\\
    &\approx 8.2\times10^7\,\text{eV} = 1.3\times10^{-11}\,\text{J}
\end{align*}

The fragmentation that defines this transition is a near-surface process: the \ce{Si+} ions that carry the spectroscopic signature originate from the outermost layer of the particle, and the energy released by the photoelectron and Auger-electron cascades is redeposited within a near-surface region of order $10$~nm, set by the photo- and Auger-electron ranges in silica at these photon energies \cite{lipp_quantifying_2022}, with core-level energy deposition localized to the nanometre scale and redistributed by the electron cascade \cite{huang_nanometer-scale_2024}. With a density $2.2$ g/cm$^3$, the corresponding shell mass is $m_\text{shell} \approx 1.6\times10^{-17}\,\text{kg}$. 

The resulting energy density is therefore on the order of
\begin{equation*}
    \frac{E_\text{abs}}{m_\text{shell}} \approx 1\,\text{MJ/kg}
\end{equation*}
of order $0.1$~eV per atom, sufficient to heat the shell to $\approx 1700$~K, near the melting point of \ce{SiO2} \cite{schweigert_structure_2002}. This threshold marks the transition from surface-limited chemistry to a warm dense matter regime, consistent with the change of slope of the \ce{Si+} feature observed experimentally.

The energy density at the transition is an intensive quantity and, at $1.88$ keV where all particles are optically thin, is independent of particle volume; the transition therefore occurs at a common absorbed fluence for every size. The detected ion count, however, is not a linear proxy for the deposited energy density: the number of ions reaching the detector saturates and does not scale with the particle volume, so the transition appears at different ion counts for different sizes, from $\approx 3{,}000$ counts for the $500$ nm particles to $\approx 2{,}000$ for the $300$ nm and $150$ nm \ce{Au}@\ce{SiO2} particles, and $\approx 1{,}500$ for the $100$ nm particles. The $500-2{,}000$ ion-count window used for the main analysis therefore lies below the transition for the larger particles and reaches it only for the smallest. This does not affect the reported field dependence: the same field ordering of the relative \ce{H+}, \ce{H2+}, and \ce{H3+} yields is recovered for windows from $500-1{,}000$ ion counts, which lies below the transition for every particle type, through $500-3{,}000$, and persists in windows above the transition (Fig.~\ref{figS10}), so the redistribution is set by the surface field rather than by the choice of intensity window.

\subsection{Surface Electric Field}

X-ray ionization induces highly charged nanoparticle surfaces through photoelectron emission. By counting the number of electrons ejected from the nanoparticle, the electrostatic field on the surface can be inferred from
\begin{equation}
    E\left(R\right)=\frac{Ne}{4\pi\epsilon_0 R^2}, \label{eq:Sfield}
\end{equation}
where $N$ is the number of emitted electrons, and $R$ is the particle radius (Fig.~\ref{figS9}(a)). The eTOF spectrometer therefore probes the surface charge directly. 

The field in Eq.~\ref{eq:Sfield} treats the ionized particle as a uniformly charged conducting sphere and neglects the dielectric response of the silica. A dielectric of relative permittivity $\varepsilon_\text{r}$ redistributes the surface charge and screens the field within the material, but the field outside the particle, which is what accelerates the emitted ions and sets the reaction environment, is fixed by the total escaped charge and the particle radius through Gauss's law and is independent of $\varepsilon_\text{r}$. Screening therefore affects the internal potential and the spatial distribution of bound charge, not the external surface field that enters our analysis. The approximation is further supported by the agreement, over their common range ($0-2$ V/nm), between the electron-counting field and the independent SIMION calibration of the \ce{Si+} kinetic energy (Fig.~\ref{figS8}), which makes no assumption about the internal dielectric. Residual deviations from this agreement bound the associated systematic below the two orders of magnitude range of fields spanned in this work.

\begin{figure}[!htb]
  \centering
  \includegraphics[width=\textwidth]{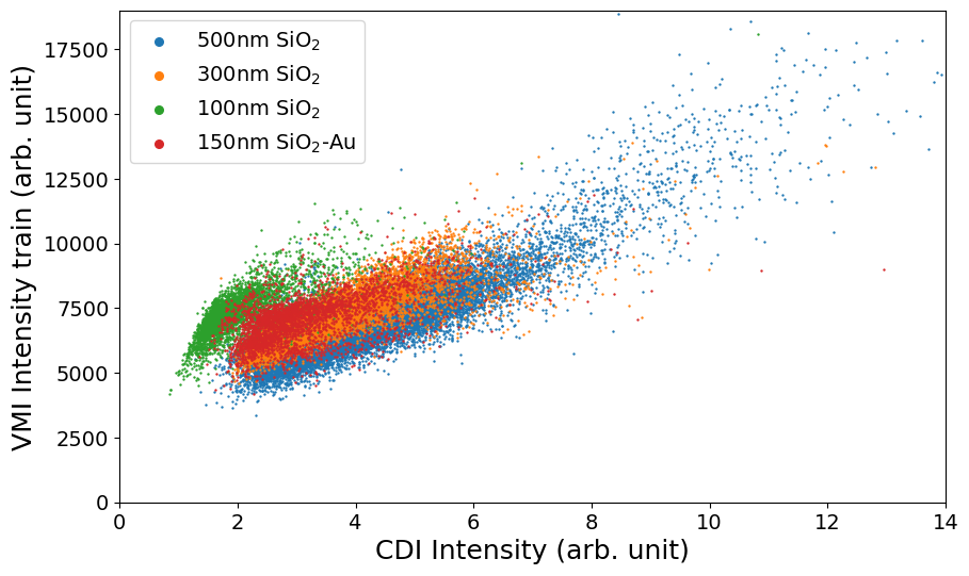}
  \caption{Intensity measured in the ion VMI corresponding to the number of ions flying into the detector, with respect to the intensity measured in the CDI corresponding to the number of scattered photons flying into the detector. The intensities are per train, which contain $92$ pulses, with a hit rate of about $1\%$. Each dot corresponds to one hit, for $500$, $300$ and $100$ nanometer diameter \ce{SiO2} and $150$ nanometer diameter core-shell \ce{Au}@\ce{SiO2} nanoparticles.}
  \label{figS7}
\end{figure}

\begin{figure}[!htb]
  \centering
  \includegraphics[width=\textwidth]{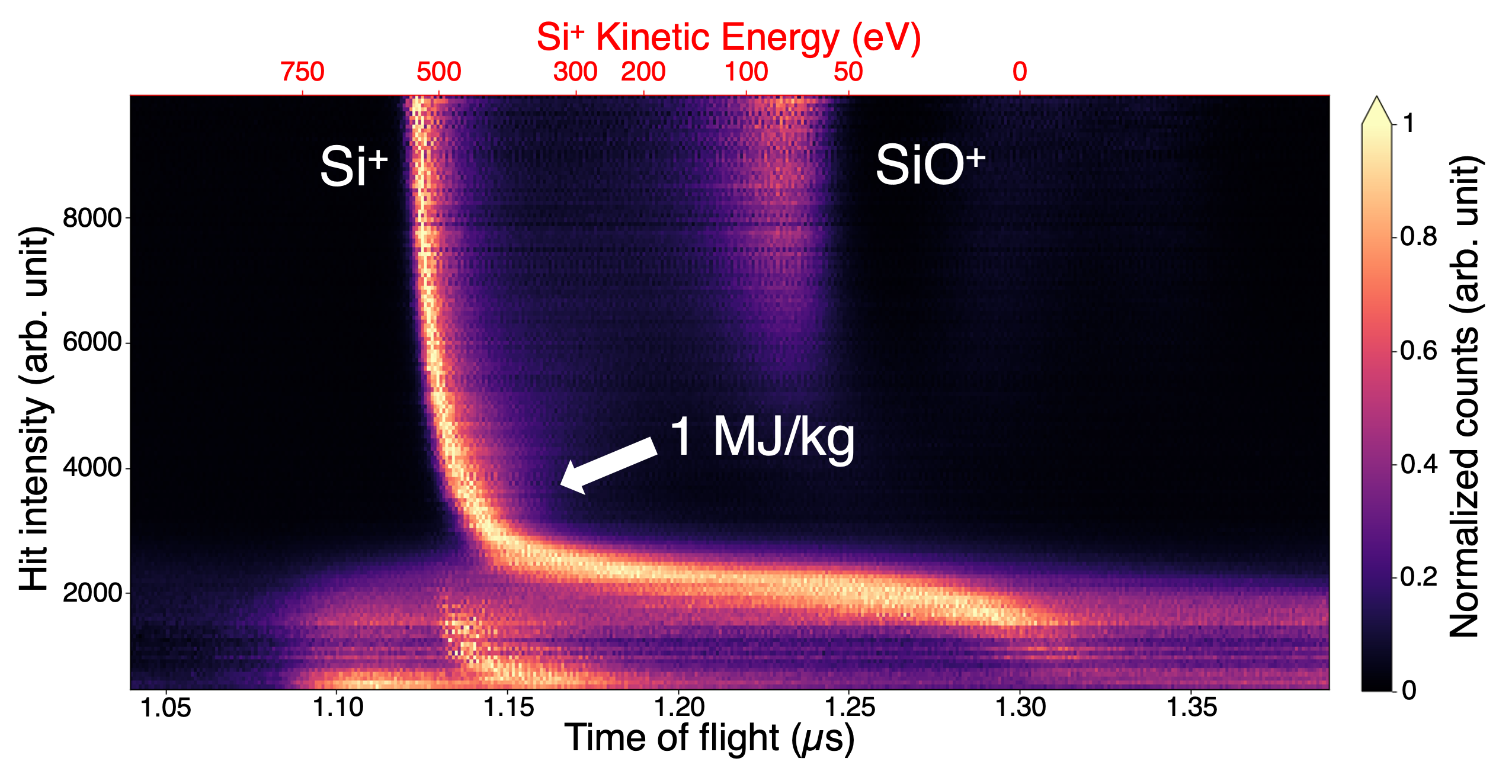}
  \caption{Silicon ion \ce{Si+} channel emitted towards the detector shown for hit intensity versus time-of-flight. Through SIMION simulations, the kinetic energy of the \ce{Si+} can be inferred. Two different regimes of reaction dynamics are separated by a change in the slope, at which an energy density of $1$ MJ/kg is reached.}
  \label{figS8}
\end{figure}

\begin{figure}[!htb]
  \centering
  \includegraphics[width=0.8\textwidth]{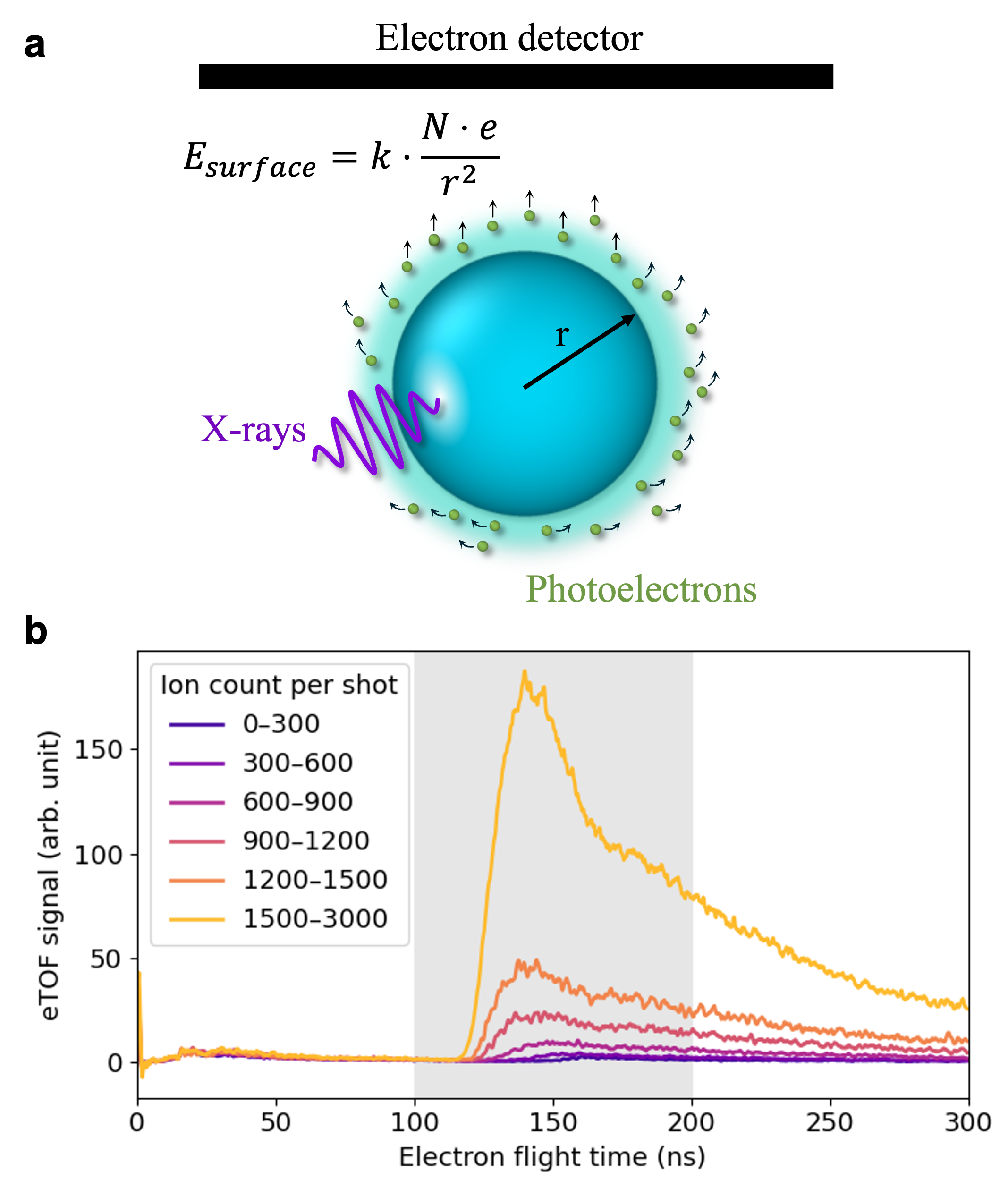}
  \caption{(a) The per-shot surface electric field $E$ is retrieved from the number of photoelectrons and the diameter of the corresponding nanoparticle. (b) Electron time-of-flight (eTOF) spectra recorded on the $300$ nm \ce{SiO2} particles, averaged over shots binned by per-shot ion count, an independent proxy for hit intensity. The photoelectron feature rises at $\approx 115$ ns and peaks near $140$ ns; the shaded band marks the $100$-$200$ ns integration window used to obtain the electron number $N$ that sets the per-shot surface field (Eq.~\ref{eq:Sfield}). The window brackets the photoelectron peak, and the yield within it grows with hit intensity. The signal decays beyond the window, so $N$ captures an intensity-dependent fraction of the emitted electrons.}
  \label{figS9}
\end{figure}

The electron number $N$ is obtained directly from the eTOF trace. For each hit, the detector is used as a counter: ions are counted over their full flight range (up to $3.54\,\mu$s), providing the per-shot intensity monitor (Fig.~\ref{figS7}), while electrons are counted in the $100-200$ ns window that contains the photoelectron feature (Fig.~\ref{figS9}(b)). The photoelectron signal rises at $\approx 115$ ns and peaks near $140$ ns, and the window brackets it. Because the escaping electrons are retarded by the Coulomb potential of the increasingly charged particle, their arrival distribution shifts to later times at higher field, so the fixed window captures a fraction of the emitted electrons that decreases with intensity, from $\approx 90\%$ at low field to $\approx 75\%$ at the highest fields reached. This introduces a systematic of at most $\approx 25\%$ in the absolute field, monotonic in intensity and therefore order-preserving, and small against the two orders of magnitude spanned by the measured fields. Structure beyond $\approx 400$ ns is backscatter off the ion-detector microchannel plate and is excluded from $N$.

The quantity measured on each shot is not the absolute electron number but the eTOF analog signal, digitized and integrated over the $100$-$200$ ns window to give a per-shot signal $S$ in digitizer units. The emitted electron number is $N = S/(g\,\eta)$, where $g$ is the single-electron response of the detector and $\eta$ the collection efficiency of the eTOF, which is below unity. Neither $g$ nor $\eta$ is calibrated independently, but both enter Eq.~\ref{eq:Sfield} only as a constant multiplicative factor and therefore rescale the entire field axis uniformly, leaving the ordering of the fields and the shapes of the field-dependent yields unchanged. The absolute field scale is instead fixed by matching the electron-counting field to the independent SIMION calibration of the \ce{Si+} kinetic energy over their common range ($0$-$2$ V/nm, Fig.~\ref{figS8}), which makes no reference to the electron signal. The collection efficiency thus need not be known absolutely: it is absorbed into the constant that this cross-check determines.

Throughout, the nominal sizes ($500$, $300$, and $100$ nm \ce{SiO2} and $150$ nm \ce{Au}@\ce{SiO2}) are used only as labels. Every field is computed from the measured TEM diameters of $504$, $312$, $126$, and $147$~nm (Fig.~\ref{figS16}). The two independent measures of the field agree over their common range. For a $500$~nm \ce{SiO2} particle at an intensity yielding $\approx 5{,}000$ detected ions, $\approx 25{,}000$ electrons are emitted, giving a surface field of $0.57$~V/nm by Eq.~\ref{eq:Sfield} and a surface potential $V_{surf} = E \cdot R = 144$ V. As a consistency check, fully converting the corresponding surface potential into ion kinetic energy yields a proton energy of $144$~eV,  which falls within the \ce{H+} distribution measured through the SIMION calibration (Fig.~\ref{fig2}(b)). The full \ce{Si+} kinetic energy range ($0-500$ eV) maps to $0-2$ V/nm for the $500$ nm particles, matching the field range electron counting reaches at the highest $500$ nm fluences. Because it is independent of the ion optics, the electron counting route is used for the per-shot field assignment.

The field rises steeply as the particle size decreases. A $300$~nm \ce{SiO2} particle at the same intensity emits $\approx 80{,}000$ electrons, corresponding to $4.7$~V/nm. A $100$~nm particle under identical illumination emits $\approx 20{,}000$ electrons, somewhat fewer than the $500$~nm particle, but its smaller radius raises the field to $7.3$~V/nm. With the measured diameters ($504$ and $126$~nm, a ratio of $\approx 4$, Fig.~\ref{figS16}) and a comparable charge, Eq. \ref{eq:Sfield} predicts roughly an order-of-magnitude increase, consistent with the inverse-square scaling and with the field rising as the particle radius decreases. Spherical core-shell \ce{Au}@\ce{SiO2} nanoparticles ($100$~nm gold core, $25$~nm silica shell) are expected to enhance the surface yield, because the gold core is an efficient soft-X-ray absorber, with a photoabsorption cross-section substantially larger than that of silica at $1.88$ keV \cite{148746}, driving a stronger photoionization and Auger cascade \cite{casta_electron_2015,lipp_quantifying_2022}, and the silica shell presents a hydroxyl-terminated, high-surface-area interface that promotes surface reactions \cite{zhuravlev_concentration_1987}; their field under identical illumination is $4.0$~V/nm.

Computing the field for each individual hit from its measured electron count and the
TEM diameter of its particle type (Eq.~\ref{eq:Sfield}, Fig.~\ref{figS16}), and binning
by field, collapses the per-type yields onto common curves (Fig.~\ref{fig3}(b)). The collapse is insensitive to the exact ion yield window used below the warm dense matter transition and the same field ordering persists in windows above it (Fig.~\ref{figS10}). The same field dependence appears in the ionized-water yield, which increases with $E$ while the field-insensitive \ce{Ar^3+} reference stays flat (Fig.~\ref{figS11}). By contrast, the silica-framework ions (\ce{Si+}, \ce{SiO+}, and the di-silicon \ce{Si_2^+} feature) show no monotonic dependence on $E$ (Fig.~\ref{figS11}), confirming that the field-controlled chemistry is decoupled from framework fragmentation.

\subsection{Yield Uncertainties and Significance}

\begin{figure}[!htb]
  \centering
  \includegraphics[width=0.8\textwidth]{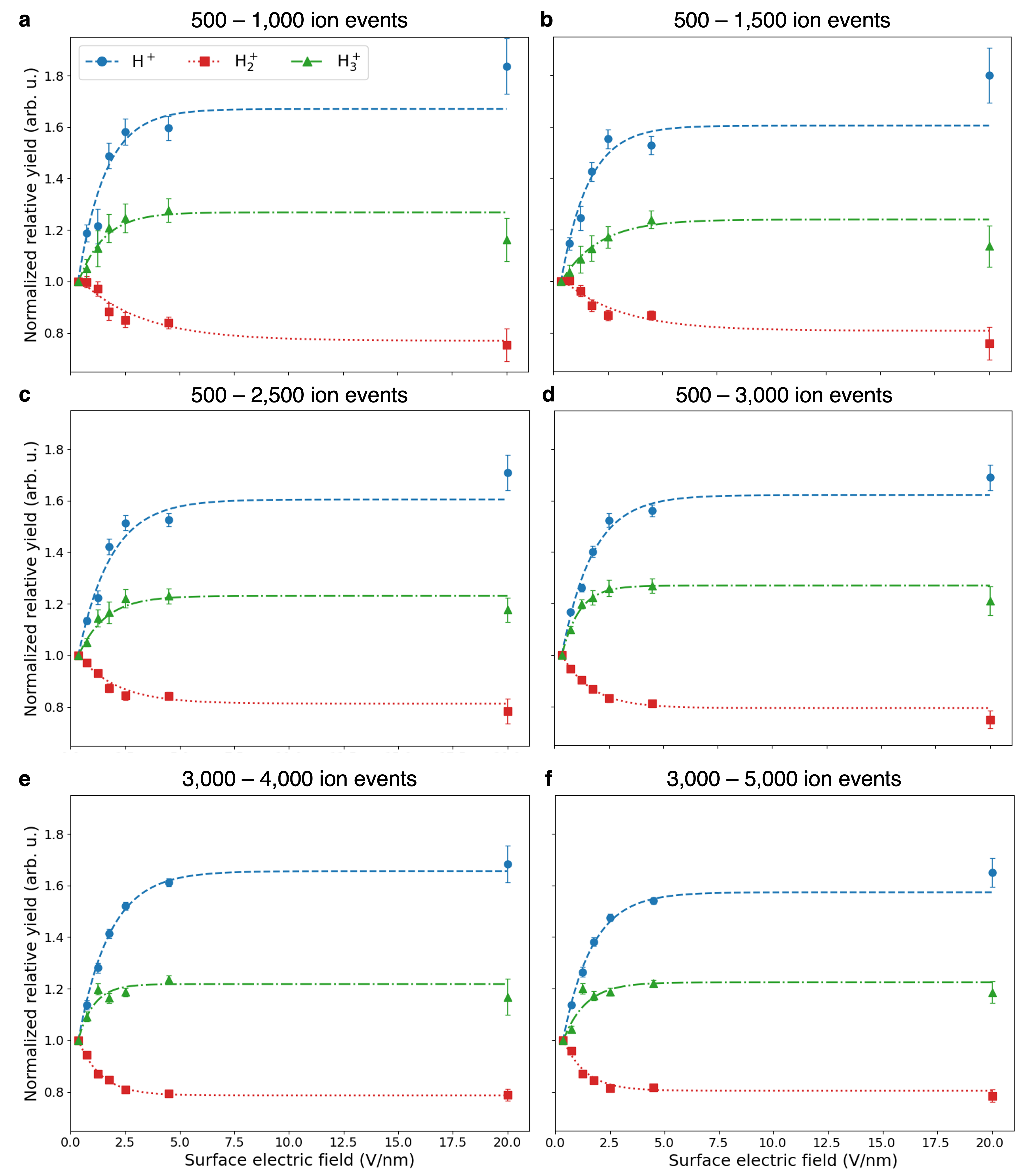}
  \caption{Relative yields of \ce{H+}, \ce{H2+}, and \ce{H3+} versus $E$, each normalized to its lowest-field value, for six hit-intensity windows: (a) $500-1{,}000$ and (b) $500-1500$ ion events, strictly below the warm dense matter (WDM) transition for all particles; (c) $500-2500$ and (d) $500-3000$ ion events, across the WDM transition; and (e) $3000-4000$ and (f) $3000-5000$ ion events, above it. The field-controlled redistribution, \ce{H+} and \ce{H3+} rising and saturating while \ce{H2+} declines, is reproduced in all six windows. Panels (a) and (b) show the result is insensitive to the exact ion yield range within the surface-chemistry regime. Panels (c)/(d) and (e)/(f) show that the field ordering of the relative yields persists respectively across and above the WDM transition, into the plasma regime where the silica framework has broken down. Error bars are $1\sigma$ shot-level bootstrap uncertainties.}
  \label{figS10}
\end{figure}

\begin{figure}[!htb]
  \centering
  \includegraphics[width=\textwidth]{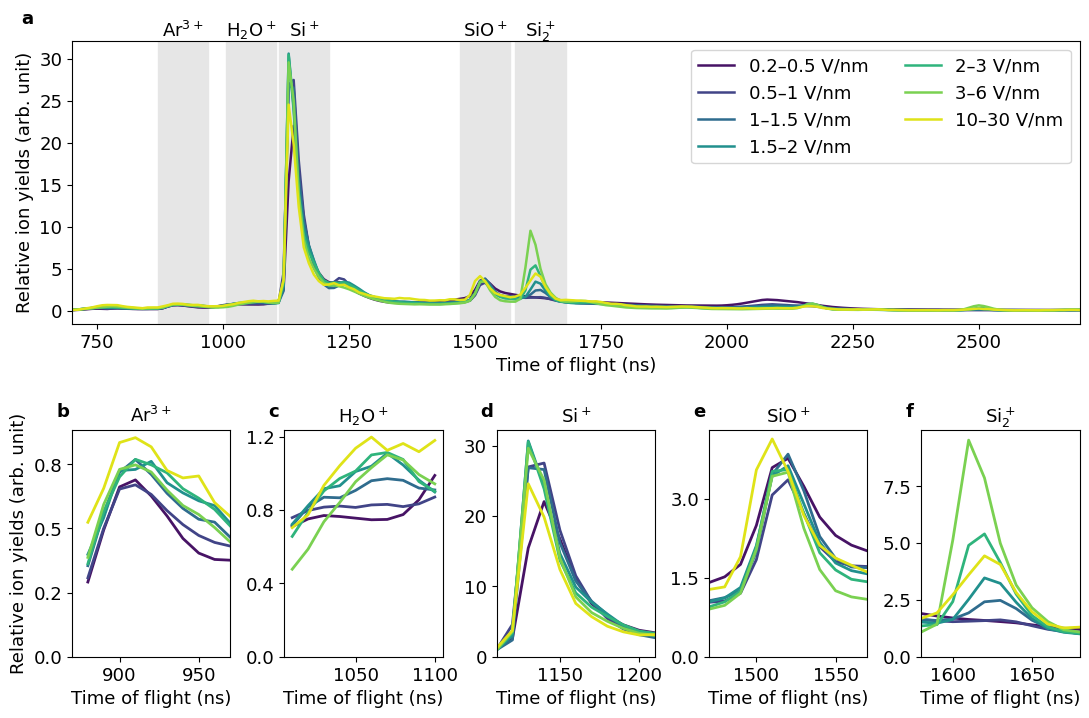}
  \caption{Field dependence of the ion time-of-flight spectrum, pooled over all particle sizes and compositions and binned by surface electric field $E$. (a) Full spectrum from $700$ to $2700$ ns; shaded bands mark the species examined in the zoomed panels. (b)-(f) Expanded views of the \ce{Ar^3+}, \ce{H2O+}, \ce{Si+}, \ce{SiO+}, and \ce{Si_2^+} peaks. The gas-phase \ce{Ar^3+} reference (b) is field-insensitive, whereas the \ce{H2O+} yield (c) rises with $E$, confirming that water ionization is governed by the local surface field. The silica-framework ions \ce{Si+} (d), \ce{SiO+} (e), and \ce{Si_2^+} (f) show no monotonic field dependence. The spectra are normalized to the total ion counts over the full time-of-flight range.}
  \label{figS11}
\end{figure}

The relative yields in Fig.~\ref{fig3}(b) and Fig.~\ref{figS10} are pooled fractions: for each field bin, $p = n_\text{peak}/n_\text{total}$, where $n_\text{peak}$ is the number of ions in the species time-of-flight window and $n_\text{total}$ the number in the full analysis window, summed over the four particle sizes and composition. The statistically independent unit is the X-ray shot, not the individual ion: ions recorded in one shot share that shot's field and composition, and the four different particles are pooled within each bin, so an ion-level binomial error $\sqrt{p(1-p)/n_\text{total}}$ underestimates the uncertainty by treating correlated ions as independent.

Uncertainties are therefore obtained by bootstrapping over shots. Within each field bin the pulse identifiers are resampled with replacement $4{,}000$ times, the pooled fraction is recomputed for each resample, and the standard deviation of the distribution is taken as the $1\sigma$ error. For yields normalized to the lowest-field bin, the reference bin is resampled in the same procedure so its uncertainty propagates into every normalized point; the reference point is unity by construction and carries no error bar.

The error model was validated on the data in four ways. (i) When the in-peak labels are randomly permuted across ions within a bin, removing all shot-level structure, the bootstrap reproduces the analytic binomial error, confirming the estimator is calibrated. (ii) Without permutation, the bootstrap exceeds the binomial by a factor of roughly $1.4$ to $2.4$, quantifying the substantial overdispersion from shot clustering and type pooling. (iii) A delete-one-shot jackknife agrees with the bootstrap to within a few percent. (iv) For the normalized yields, the bootstrap matches first-order ratio propagation, $\sigma_R = R\sqrt{(\sigma_i/p_i)^2 + (\sigma_0/p_0)^2}$, to one percent. The bootstrap error was confirmed converged in the number of resamples.

The field dependence of each species was assessed by fitting the saturating (\ce{H+}, \ce{H3+}) or decaying (\ce{H2+}) form weighted by these errors. The amplitudes differ from zero at $28.9\sigma$ (\ce{H+}), $10.2\sigma$ (\ce{H2+}), and $8.7\sigma$ (\ce{H3+}). A likelihood-ratio comparison of the sloped fit against a constant gives $p < 10^{-3}$ for each species, establishing that the surface field governs all three hydrogen ion yields.

\subsection{Morphology Classification by Convolutional Neural Network}

A convolutional neural network (CNN) was trained to classify single shot diffraction patterns as monomer or dimer cases. Inputs are $200\times200\times2$ tensors comprising the preprocessed scattering image from the pn-junction Charge Coupled Device (pnCCD) detector and a binary beam-stop mask. Preprocessing mirrors each raw frame to a square field while preserving symmetry and dynamic range (normalized to $[0,1]$), and a fixed beam-stop mask ensures the network learns only from informative pixels. The architecture is a masked CNN: a $7\times7$ convolutional layer ($16$ filters) with batch normalization, ReLU (rectified linear unit) activation function, element-wise masking, and $4\times4$ max-pooling. A single residual block combines a $1\times1$ skip connection with a $5\times5$ convolutional layer ($32$ filters), batch normalization, ReLU activation, and spatial dropout. The resulting feature map is reduced by global average pooling and connected to a dense layer of $64$ neurons (ReLU activation, L2 regularization) before the final sigmoid output layer. Optimization was performed using the Adam optimizer (learning rate $10^{-4}$) and a binary focal loss with $\alpha=0.8$ and $\gamma=2.0$, with accuracy, recall, and precision tracked as metrics. The data were split into $80\%$ training and $20\%$ testing sets with stratification and class weighing to account for class imbalance. The model trained with a batch size of $32$ for up to $40$ epochs, using a \texttt{ReduceLROnPlateau} callback (factor $0.5$, patience $5$) and \texttt{EarlyStopping} (patience $10$, restoring best weights).

A progressive learning scheme was employed, because of the difficulty of classifying low-intensity shots, even to the human eye. The model was trained five times with decreasing learning rates, each iteration including data with increasingly lower hit intensities to first teach the network to recognize high-signal images and then fine-tune it with more challenging, low-intensity shots. The scarcity of the dataset led to augmentation techniques, generating a left-right flipped version and four random intensity-scaled copies with scaling factors drawn uniformly from the range $0.4-2.5$. To further balance the dataset, as dimers were underrepresented in the data, additional synthetic dimer diffraction patterns were generated. Each synthetic image represents the coherent sum of the far-field intensities from two spherical scatterers with parameters randomly drawn from distributions matching the experimental nanoparticle size range and separation statistics. The relative center-to-center distance, azimuthal orientation, and polarization angle were randomized within physically realistic bounds to reproduce the fringe spacing and contrast observed experimentally. Intensity normalization and additive Gaussian noise were applied to mimic shot-to-shot photon fluctuations and detector noise. The simulated diffraction patterns were subsequently rescaled and masked using the same preprocessing pipeline as the real data, ensuring consistent input dimensionality and dynamic range for CNN training. These synthetic dimers were used exclusively for training and validation, enabling the network to learn generalized scattering features even where experimental dimer statistics were sparse.

Model evaluation employed dual decision thresholds ($0.35$ for monomer, $0.65$ for dimer) to reject ambiguous cases. The resulting confusion matrix (Fig.~\ref{figS13}) gives $TN=1062$, $FP=1$, $FN=5$, and $TP=457$ for a total of $N=1525$ classified shots. Treating dimers as the positive class, the model reaches a precision of $457/(457+1)=0.998$, a recall of $457/(457+5)=0.989$, and an F1-score of $0.993$. The CNN thus distinguishes monomeric from dimeric diffraction patterns with high fidelity, misclassifying only six events in total, providing the monomer/dimer labels used for the patterns in Fig.~\ref{fig3}(c) and the hydrogen ion yields comparison in Fig.~\ref{fig3}(d).

\subsection{Scattering Images Analysis}

Real-space length scales are extracted from the fringe spacings in cropped regions of the pnCCD detector. The full detector has \(1024\times1024\) pixels with a pixel size of \(75\times75~\mu\mathrm{m}^2\). For the patterns in Fig.~\ref{fig3}(c), a $200\times100$ pixel region centered on the forward-scattering direction is selected, containing the main ring structure while excluding the masked direct-beam region.

The X-ray photon energy is \(E = 1.88~\mathrm{keV}\), which corresponds to a wavelength of
\begin{equation}
    \lambda = \frac{hc}{E} = \frac{1239.84~\mathrm{eV\,nm}}{1880~\mathrm{eV}} \approx 0.66~\mathrm{nm}
\end{equation}
where \(h\) is Planck's constant and \(c\) is the speed of light. The sample-detector distance is \(L = 346~\mathrm{mm}\). For small scattering angles \(\theta\), the scattering geometry is well approximated by
\begin{equation}
\theta = \arctan\!\left(\frac{y}{L}\right) \approx \frac{y}{L},
\end{equation}
where \(y\) is the distance on the detector from the forward direction.

The size of a scattering sphere can be determined from the spacing of the interference fringes in the angular domain. In the framework of Mie scattering and simple interferometry, the sphere diameter \(D\) is expressed as
\begin{equation}
D =\frac{m\lambda}{\sin\theta},
\end{equation}
where \(m\) is the fringe order. Using the small-angle approximation \(\sin\theta \approx \theta \approx y/L\), this relation can be written as
\begin{equation}
D \approx \frac{m \lambda L}{y}.
\end{equation}

In the top panel of Fig.~\ref{fig3}(c), corresponding to scattering from a single nanoparticle, we determine the fringe spacing by measuring the distance between four consecutive rings in the radial direction. The distance between these four rings is \(38\) pixels, which corresponds to a physical distance on the detector of
\[
y_{\mathrm{single}} = 38 \times 75~\mu\mathrm{m} = 2.85~\mathrm{mm}.
\]
Taking \(m=4\) for these four fringe spacings and inserting the experimental parameters into the expression above yields
\[
D_{\mathrm{NP}} \approx \frac{4 \,\lambda\, L}{y_{\mathrm{single}}}
                 = \frac{4 \times 0.66~\mathrm{nm} \times 346~\mathrm{mm}}{2.85~\mathrm{mm}}
                 \approx 320~\mathrm{nm}.
\]
Thus, the single nanoparticle in the top panel of Fig.~\ref{fig3}(c) has an effective diameter of approximately \(D_{\mathrm{NP}} \approx 320~\mathrm{nm}\), consistent with the expected particle size.

The bottom panel of Fig.~\ref{fig3}(c) shows a scattering pattern acquired when two nanoparticles are simultaneously in the focus, giving rise to an additional interference modulation along the ring. In this case, the fringe pattern along the ring encodes the center-to-center separation \(s\) between the two nanoparticles. In the far-field, the intensity can be written as
\begin{equation}
    I(\mathbf{q}) \propto |F(\mathbf{q})|^2[2+2\cos(\mathbf{q}\cdot\mathbf{s})],
\end{equation}
so that the fringe period in angle satisfies an analogous relation \(s = m\lambda/\sin\theta\). Using the same small-angle approximation, we obtain
\begin{equation}
    s \approx \frac{m \lambda L}{y_{\mathrm{dimer}}},
\end{equation}
where \(y_{\mathrm{dimer}}\) is now the distance on the detector corresponding to \(m\) fringes along the ring. From the bottom panel of Fig.~\ref{fig3}(c) we measure that five fringes span a distance of \(50\) pixels along the ring, which corresponds to
\[
y_{\mathrm{dimer}} = 50 \times 75~\mu\mathrm{m} = 3.75~\mathrm{mm}.
\]
With \(m=5\), this yields
\[
s_{meas} \approx \frac{5 \,\lambda\, L}{y_{\mathrm{dimer}}}
       = \frac{5 \times 0.66~\mathrm{nm} \times 346~\mathrm{mm}}{3.75~\mathrm{mm}}
       \approx 305~\mathrm{nm}.
\]
Hence, the two nanoparticles contributing to the lower scattering image in Fig.~\ref{fig3}(c) are separated by a center-to-center distance of approximately \(s_{meas} \approx 305~\mathrm{nm}\). 

Because only the component of $\mathbf{s}$ lying in the detector plane enters the azimuthal interference term, $s_\text{meas}$ is the in-plane projection of the true three-dimensional separation. For a dimer axis at angle $\alpha$ to the beam (detector normal), the projected separation is $s_\text{meas} = s\,\sin\alpha$, and for randomly oriented dimers the in-plane projection of a uniformly distributed axis averages to $\langle\sin\alpha\rangle = \pi/4$. The typical three-dimensional separation is therefore
\begin{equation*}
    \langle s\rangle = \frac{\langle s_\text{meas}\rangle}{\langle\sin\alpha\rangle} = \frac{4}{\pi}\,\langle s_\text{meas}\rangle \approx 1.27\,\langle s_\text{meas}\rangle \approx 390~\text{nm},
\end{equation*}
corresponding to a typical edge-to-edge gap of $G = \langle s\rangle - D_\text{NP} \approx 70$~nm. The per-event field superposition described next uses the spacing measured for each individual dimer rather than this orientation-averaged value.

\subsection{Dimer Surface Field by Superposition}

For the monomer/dimer comparison, the field is referenced at a fixed detected ion count and the corresponding electron count is read per nanoparticle. A $300$~nm \ce{SiO2} monomer at $\approx 4{,}000$ detected ions emits $\approx 54{,}000$ electrons, giving a surface field of $3.2$~V/nm. A dimer at $\approx 8{,}000$ detected ions emits $\approx 97{,}000$ electrons, or $\approx 48{,}500$ per particle, giving $2.9$~V/nm each. The neighbor's field is then added by linear superposition at the interparticle separation read from the two-particle interference fringes: with photon wavelength $\lambda = 0.66$~nm and detector distance $L = 346$~mm, a center-to-center separation $d$ produces fringes of spacing $\lambda L/d$, resolved from $\approx 3$ pixels per fringe at $d \approx 1\,\mu$m down to $\approx 10$ pixels per fringe at contact ($d \approx 300$~nm for two $300$~nm spheres). Contact pairs are therefore within the resolved range. This superposition treats the particles as independent point-charge sources and neglects mutual shadowing, dielectric screening, and the altered geometry of fused or necked aggregates, which do not present as two-sphere diffraction patterns.

Selecting monomer and dimer events at matched effective field ($2$-$4$~V/nm; other ranges in Fig.~\ref{figS14}) yields overlapping hydrogen ion TOF spectra across the full resolved separation range, including contact (Fig.~\ref{fig3}(d)), demonstrating that the chemistry is governed by the local surface field alone. Aggregation enters solely through its contribution to that field, not through any additional chemistry.

\subsection{Nonadiabatic Quantum Molecular Dynamics Simulations}

Nonadiabatic quantum molecular dynamics (NAQMD) simulations were performed for both single- and double-hole generation within the bulk of the silica slab. The average hole population, resolved between the slab and the adsorbed water layer, transfers from the slab into the water within $\approx 60$ fs (Fig.~\ref{figS15}(a)). Following this transfer the field-driven charge in the water layer fragments the surface water molecules. The simulation snapshots in Fig.~\ref{figS15}(b) track one such event, in which the highlighted oxygen leaves the surface as part of an ejected \ce{H2O+} between $90$ and $110$ fs, while an example \ce{H3O+} formed in the same process remains confined within the hydrogen-bond network and is slower to desorb. Depending on the charge state and clustering of the water molecules, many fragmentation pathways occur \cite{furuhama_reactions_2006}, with \ce{H2+} among the possible channels.

\clearpage

\begin{figure}[!htb]
  \centering
  \includegraphics[width=\textwidth]{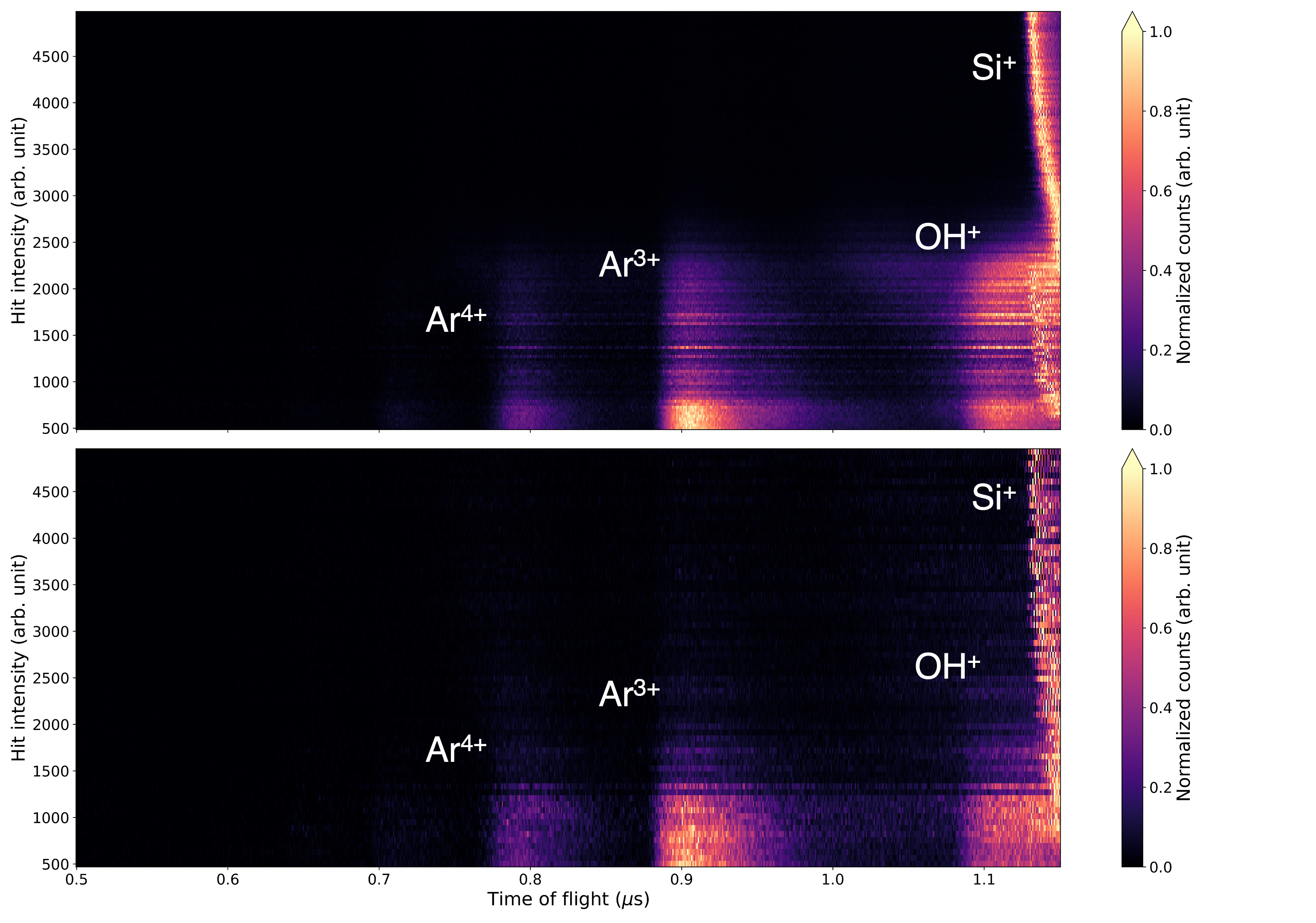}
  \caption{Hit intensity versus time-of-flight showing the absence of \ce{C+}, which would show between \ce{Ar^4+} and \ce{Ar^3+}, and \ce{CH3+}, with an expected peak around $1 \mu$s, ruling out ethoxy groups from the synthesis, or adventitious surface hydrocarbons, as the trihydrogen source. Top shows the commercial $500$ nm particles, and bottom the TEOS grown $100$ nm particles.}
  \label{figS12}
\end{figure}

\begin{figure}[!htb]
  \centering
  \includegraphics[width=0.4\textwidth]{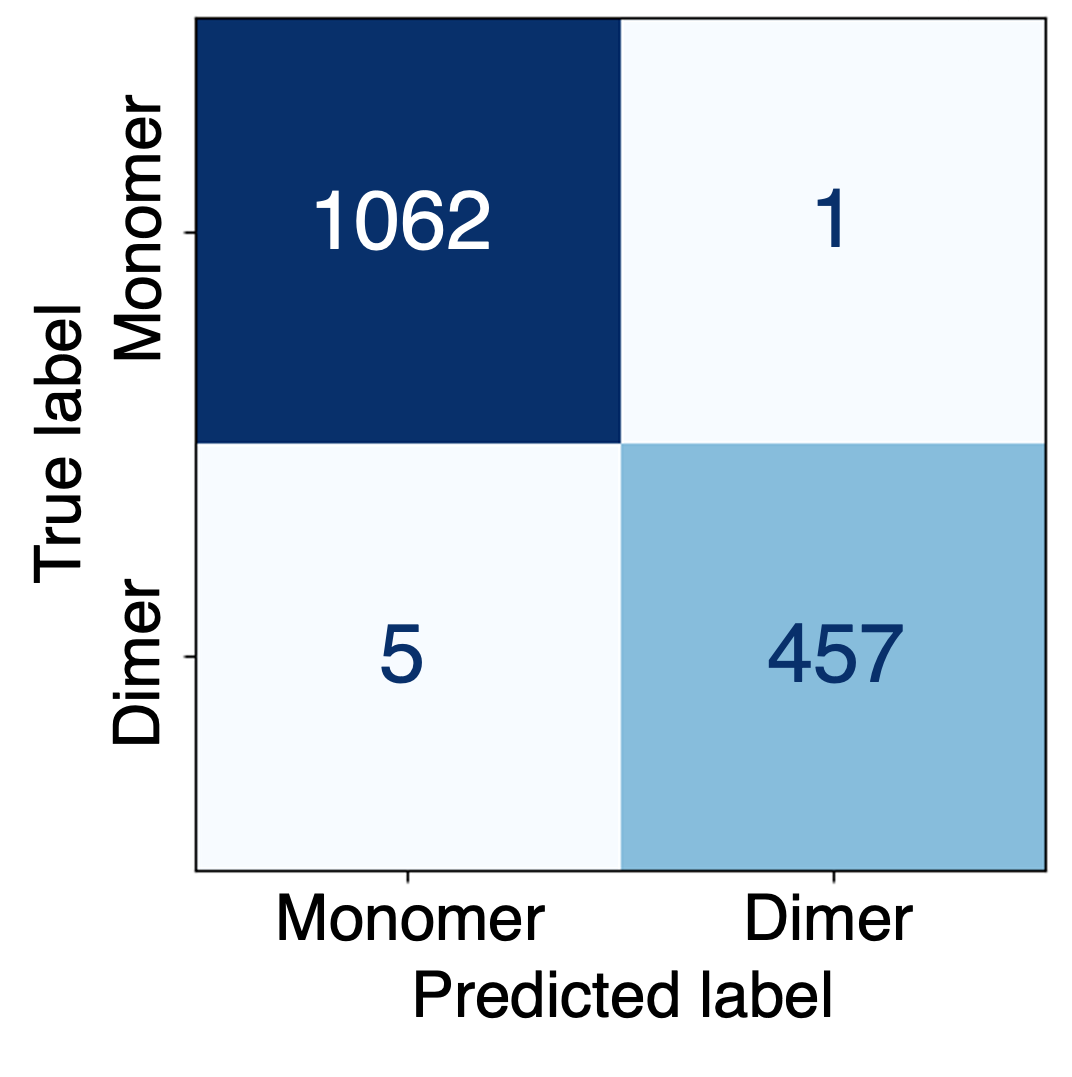}
  \caption{Confusion matrix obtained using dual classification thresholds ($0.35$ and $0.65$)
for monomer and dimer nanoparticle diffraction patterns. The convolutional neural network (CNN) achieves discrimination with an F1-score $\geq 99\%$.}
  \label{figS13}
\end{figure}

\begin{figure}[!htb]
  \centering
  \includegraphics[width=\textwidth]{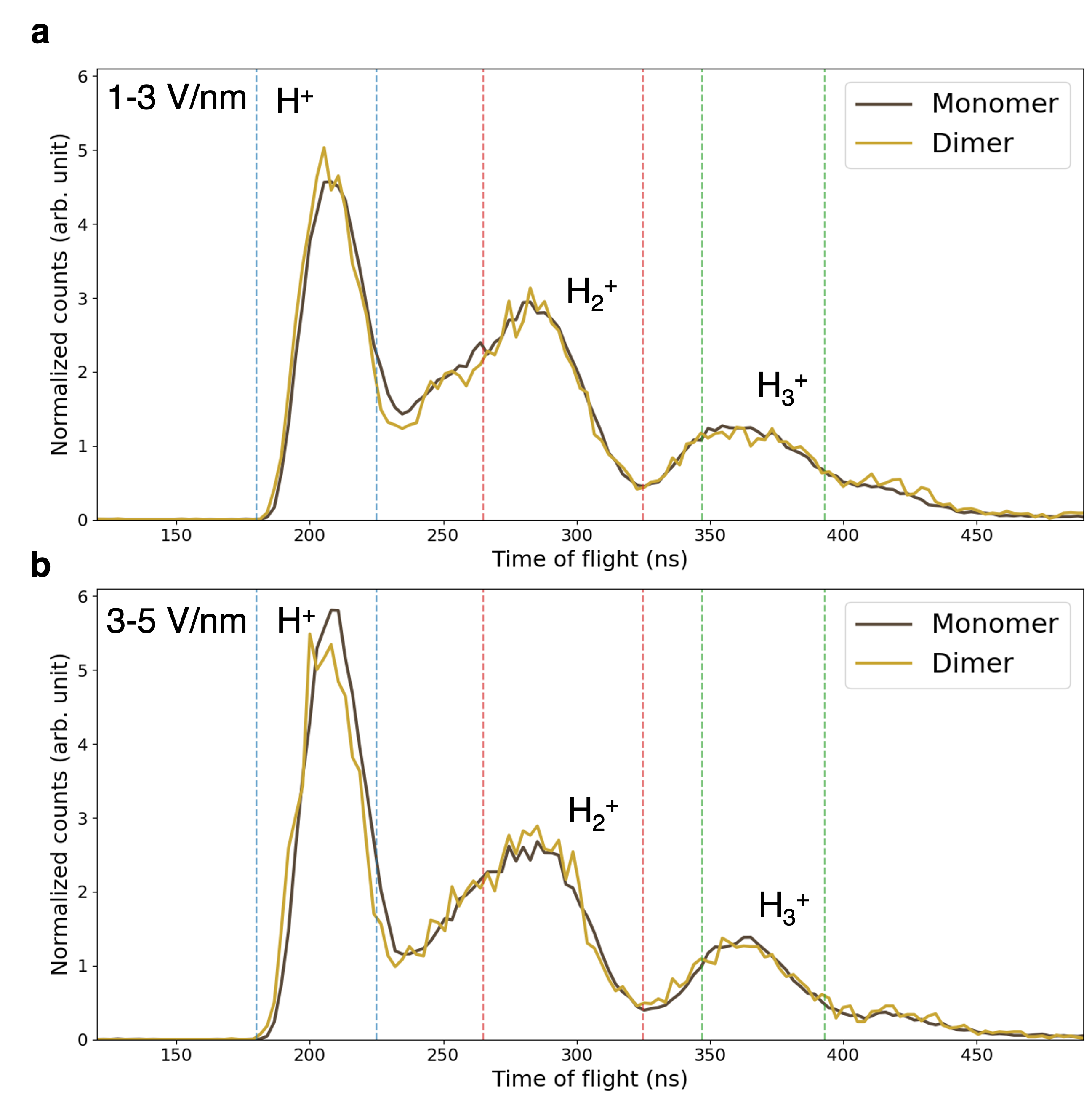}
  \caption{Time-of-flight spectra of monomer and dimer events selected at different effective field ranges: (a) $1-3$ V/nm and (b) $3-5$ V/nm.}
  \label{figS14}
\end{figure}

\begin{figure}[!htb]
  \centering
  \includegraphics[width=0.7\textwidth]{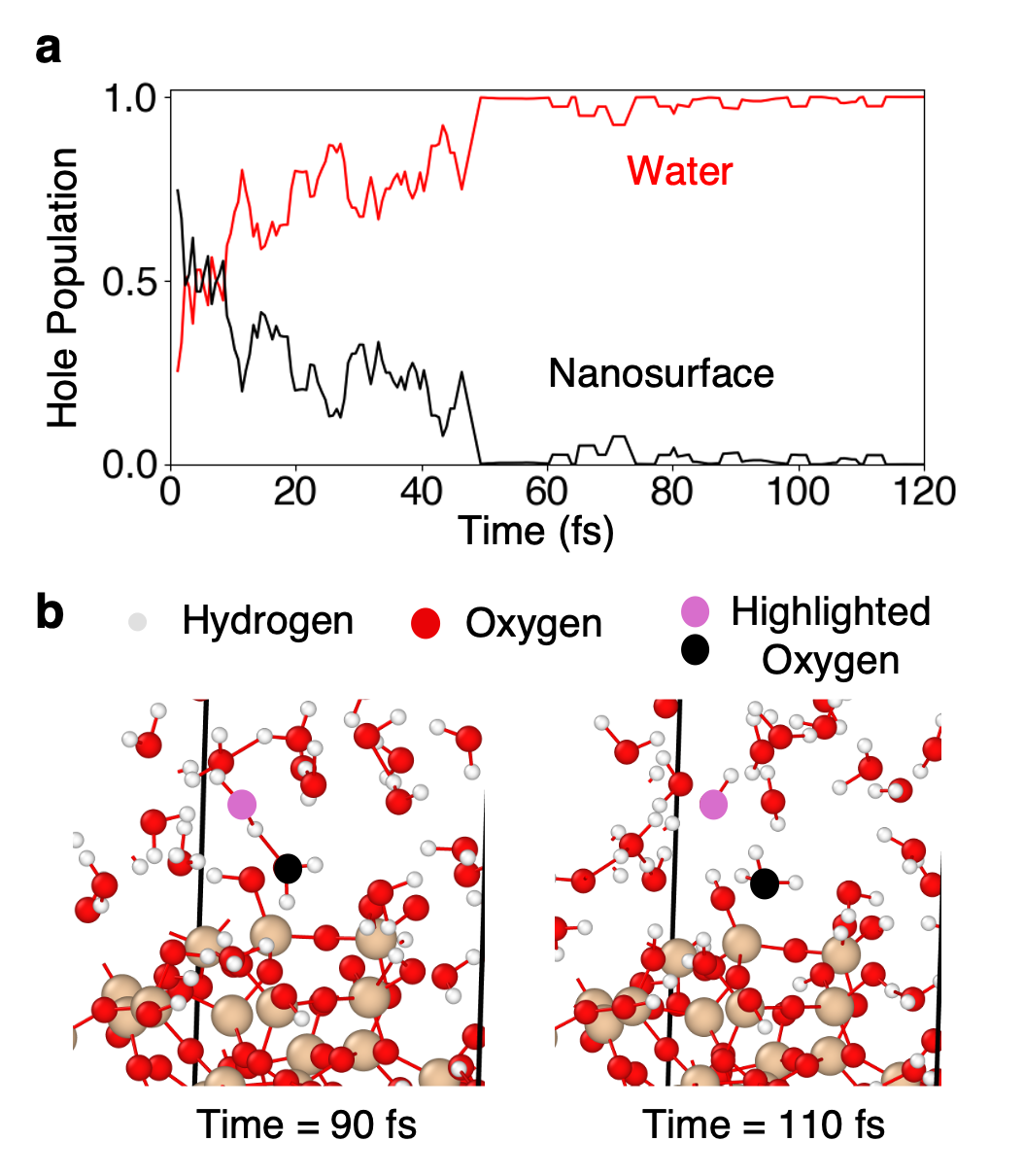}
  \caption{Nonadiabatic quantum molecular dynamics (NAQMD) simulations of the hydrated silica surface at $5$ V/nm. (a) Time evolution of the average hole population on the silica (\ce{SiO2}) slab and in the adsorbed water layer. The hole is transferred from the slab into the water layer within $\approx 60$ fs. (b) Simulation snapshots after the transfer: the field-driven charge fragments the water molecules and \ce{H2O+} leaves the surface, while an example \ce{H3O+} formed in the same process remains confined within the hydrogen-bond network and desorbs more slowly.}
  \label{figS15}
\end{figure}

\begin{figure}[!htb]
  \centering
  \includegraphics[width=\textwidth]{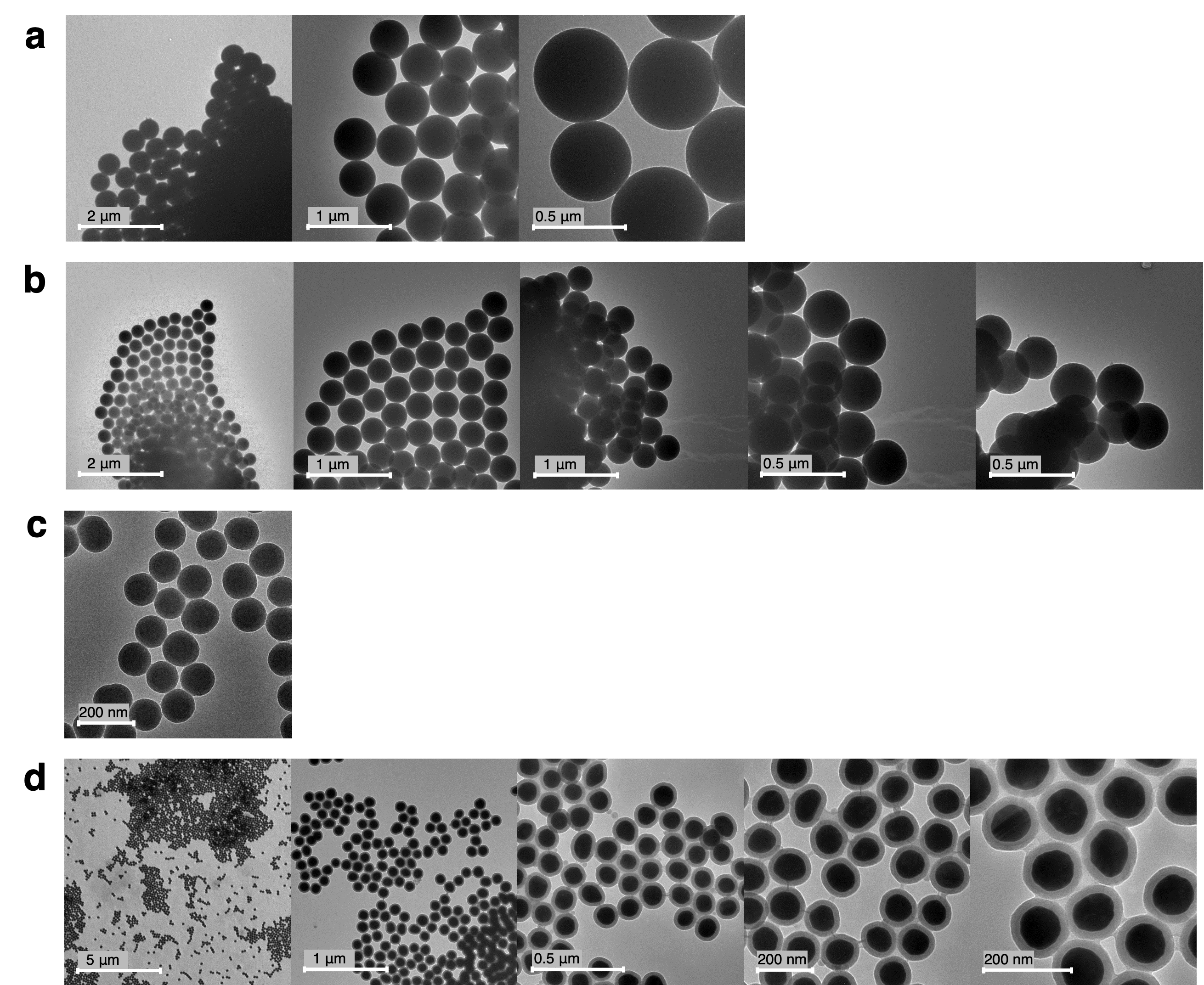}
  \caption{Transmission electron microscopy (TEM) images of the (a) $500$ nm \ce{SiO2} nanoparticles of $10$ g/L concentration in a solution of \ce{H2O} with measured diameter $504 \pm 32$ nm; (b) $300$ nm \ce{SiO2} nanoparticles of $10$ g/L concentration in a solution of \ce{H2O} with measured diameter $312 \pm 14$ nm;  (c) $100$ nm \ce{SiO2} nanoparticles of $10$ g/L concentration in a solution of \ce{H2O} with measured diameter $126 \pm 7$ nm; (d) $150$ nm \ce{Au}@\ce{SiO2} core-shell nanoparticles of $1$ g/L concentration in a solution of \ce{H2O} with measured diameters of the core $100 \pm 7$ nm and of the core-shell $147 \pm 9$ nm.}
  \label{figS16}
\end{figure}

\end{document}